\documentstyle[12pt,equations,cite]{article}
\def\bold#1{\setbox0=\hbox{$#1$}%
     \kern-.025em\copy0\kern-\wd0
     \kern.05em\copy0\kern-\wd0
     \kern-.025em\raise.0433em\box0 }

\def\slash#1{\setbox0=\hbox{$#1$}#1\hskip-\wd0\dimen0=5pt\advance
       \dimen0 by-\ht0\advance\dimen0 by\dp0\lower0.5\dimen0\hbox
         to\wd0{\hss\sl/\/\hss}}
\newcommand{\dd}{\displaystyle}
\newcommand{\nn}{\nonumber}

\def\Re{\mathop{{\cal R}\!e}}
\newcommand{\smallz}{{\scriptscriptstyle Z}} 
\newcommand{\smallw}{{\scriptscriptstyle W}} %
\newcommand{\smallh}{{\scriptscriptstyle H}} %
\newcommand{\eps}{\epsilon}

\newcommand{\chat}{ {\hat c} }
\newcommand{\acur}{ {\hat \alpha} }
\newcommand{\mz}{m_\smallz}
\newcommand{\mw}{m_\smallw}
\newcommand{\mh}{m_\smallh}
\newcommand{\mt}{m_t}

\newcommand{\mwb}{m_{\smallw_0}}

\newcommand{\scur}{\mbox{$\hat{s}^2$}}
\newcommand{\sincur}{\mbox{$\sin^{2}\!\hat{\theta}_{\scriptscriptstyle W} 
                           (\mz^2)$}}
\newcommand{\ccur}{\mbox{$\hat{c}^2$}}

\newcommand{\dr}{\mbox{$ \Delta r$}}
\newcommand{\de}{\mbox{$ \delta e^2$}}
\newcommand{\ds}{\mbox{$ \delta s^2$}}
\newcommand{\deoe}{ \frac{\delta e^2}{e^2}}
\newcommand{\dsos}{ \frac{\delta s^2}{s^2}}
\newcommand{\dmw}{\mbox{$ \delta \mw^2$}}
\newcommand{\drcar}{\mbox{$\Delta \hat{r}$}}

\newcommand{\azz}{\mbox{$ A_{\smallz \smallz} $}}
\newcommand{\aww}{\mbox{$ A_{\smallw \smallw} $}}

\newcommand{\gmu}{\mbox{$ G_\mu $}}

\newcommand{\amtd}{\mbox{$ O(\alpha^2 m_t^2 /\mw^2) $}}
\newcommand{\amtq}{\mbox{$ O(\alpha^2 m_t^4 /\mw^4) $}}
\newcommand{\gmuq}{\mbox{$ O(G_\mu^2 m_t^4) $}}

\newcommand{\self}{self-energy}

\newcommand{\ew}{electroweak}
\newcommand{\msbar}{\overline{MS}}
\def\lequiv{\raise 0.4ex \hbox{$<$} \kern -0.8 em \lower 0.62 ex \hbox{$\sim$}}
\def\gequiv{\raise 0.4ex \hbox{$>$} \kern -0.7 em \lower 0.62 ex \hbox{$\sim$}}

\newcommand{\Vw}{\mbox{$V_\smallw$}}
\newcommand{\Bw}{\mbox{$B_\smallw$}}

\newcommand{\equ}[1]{Eq.~(\ref{#1})}
\newcommand{\eqs}[1]{Eqs.~(\ref{#1})}

\newcommand{\efe}[1]{Ref.\cite{#1}}

\newcommand{\be}{\begin{equation}}
\newcommand{\ee}{\end{equation}}
\newcommand{\een}{\end{subequations}}
\newcommand{\ben}{\begin{subequations}}
\newcommand{\beq}{\begin{eqalignno}}
\newcommand{\eeq}{\end{eqalignno}}
\newcommand{\beqtwo}{\begin{eqaligntwo}}
\newcommand{\eeqtwo}{\end{eqaligntwo}}
\newcommand{\beqs}{\begin{eqalignno*}}
\newcommand{\eeqs}{\end{eqalignno*}}
\newcommand{\bea}{\begin{eqnarray}}
\newcommand{\eea}{\end{eqnarray}}

\renewcommand{\thefootnote}{\fnsymbol{footnote} }
\begin{document}

\begin{titlepage}

\begin{flushright}
        \small
        hep-ph/9603374\\
        DFPD 96/TH/20\\
        MPI-PhT-96-17 
\end{flushright}
\vspace{1.2cm} 
\begin{center}
{\Large\bf 
Two-loop heavy top effects   \\
           on the $\mz$--$\mw$ interdependence }
 \\
\vspace{2cm}
{\sc Giuseppe Degrassi$^a$, Paolo Gambino$^b$, and Alessandro Vicini$^a$}\\
\vspace{.4cm}
{\em $^a$ Dipartimento di Fisica, Universit\`a  di Padova,
Sezione INFN di Padova \\
Via Marzolo 8, 35131 Padova, Italy} \\ 
\vspace{.3cm}
{\em $^b$ Max Planck Institut f\"ur Physik, Werner Heisenberg Institut,\\
 F\"ohringer Ring 6, D80805 M\"unchen, Germany}
\end{center}
\vspace{2cm} 
\begin{center}
{\bf Abstract}
\end{center}
\vspace{0.2cm}
The $\amtd$ correction to  the relation between 
$G_\mu$ and the vector boson masses is computed in the $\msbar $
scheme, and the results are used to investigate the magnitude of the effect
on the theoretical prediction of $\mw$ and $\sincur$ from $\alpha$, $G_\mu$,
and $\mz$.

\vfill\nopagebreak
\end{titlepage}
\setcounter{footnote}{0}
\renewcommand{\thefootnote}{\arabic{footnote}}

The interdependence between $\mw$, $\mz$, and $G_\mu$ has  been studied 
for a long time.
The original one-loop calculation of \dr\ \cite{Si80} has subsequently
been augmented by  the inclusion of higher order
 corrections related to
mass singularity contributions and heavy  top effects. 
The inclusion of the leading logarithms of 
$O(\alpha \ln(\mz/m_f))^n$
(here $m_f$ is a generic fermion mass) in \dr\ was  investigated in 
Refs. \cite{AM,Si84}, while Consoli, Hollik, and Jegerlehner \cite{CHJ} showed
how to take into account, in the On-Shell (OS) scheme, the  leading 
two-loop contribution
of a heavy top, namely the term that scales as $\mt^4$. A similar analysis
concerning the leading top-mass power correction 
was performed  in the $\msbar$ 
framework in \efe{DFS}. More recently, the full Higgs dependence of the 
leading $\mt^4$ contribution
was calculated by several groups \cite{bar,DFG}.
Our knowledge of mass singularity contributions 
to  \dr\ goes actually  beyond
the two-loop leading effects, as   the incorporation of
  the $O(\alpha^2\ln(\mz / m_f))$ terms was presented in \efe{Si84}.
Concerning the two-loop top corrections, however,  a
discussion of the \amtd\ correction is still missing.

Indeed, the uncertainty coming from the unknown higher order contributions
can be ascribed mainly to our ignorance of the 
$O(\alpha^2 \mt^2/\mw^2)$, as  
two and three-loop QCD corrections seem to be well under control
\cite{sirYB}, and two-loop heavy Higgs effects 
have been shown to be negligible
\cite{Velt}. A first investigation of the potential magnitude of 
higher order corrections of \ew\ origin was carried out
 by the Working Group on Precision Calculations (WGPC) at CERN \cite{YB}. 
The results of five different computer codes for the evaluation of radiative 
corrections were compared; the codes were based on different renormalization
frameworks, and  allowed various resummation options, all equivalent 
at the order of known contributions, and differing precisely at
$O(\alpha^2 \mt^2/\mw^2)$. Therefore, in  comparing the results obtained by 
choosing different options, one could have an indication 
on the importance of the 
higher-order corrections which have not yet been explicitly calculated.
Although in most cases the experimental precision is well above the 
uncertainty obtained in this way, 
one of the conclusions of the WGPC report was  that a full calculation
of $O(\alpha^2 \mt^2/\mw^2)$ would greatly  reduce the theoretical error
originating from higher order effects.

In this respect, it is interesting to note that 
the estimate of the WGPC for the theoretical uncertainty on observables 
like $\mw$ and the effective mixing angle measured at LEP is in very good
 agreement with \efe{us}, where the result of a complete calculation 
of the $O(\alpha^2 \mt^2/\mw^2)$ effects in the $\rho$ parameter for 
$\nu_\mu-e$ scattering was used as the basis for an extrapolation to 
\dr. In general,  the result was that  $O(\alpha^2 \mt^2/\mw^2)$
could be as large as the leading $\mt^4$ contributions.
In particular, the estimate of \efe{us} was that neglected \amtd\
effects could shift the theoretical prediction of the mass of the W boson
by up to about 23 MeV, depending on the Higgs boson and top masses. More 
interestingly, the prediction of the effective sine measured on the Z peak 
at LEP and SLC, $\sin^2 \theta_{eff}^{lept}$, could be shifted by up to 
1.4$\times 10^{-4}$, closer to the present experimental accuracy of 
3$\times 10^{-4}$.
Also in view of the prospects of improving the experimental accuracy 
on these observables, we feel that
a complete calculation of this kind of effects on the interdependence 
between $\mw$, $\mz$, and $G_\mu$ is  indeed timely. 
 It is therefore the aim of this paper to provide 
explicit analytical expressions for the \amtd\ contributions to \dr\
and to investigate their magnitude. 

To begin our discussion of the electroweak
corrections of \amtd\ to \dr\ 
we write the relation between the $\mu$-decay constant and the 
charged current amplitude expressed in terms of bare quantities. At the
two-loop level  we have 
\ben \label{e2.1}
\beq
\frac{\gmu}{\sqrt2} = & \frac{e_0^2}{8 s_0^2\,\mwb^2} \left\{1+ \dr_0\right\}
\label{e2.1a}\\
 \dr_0=& \left\{
 -\frac{\aww(0)}{\mwb^2} + \Vw + \mwb^2 \Bw +
\dd\frac{\aww(0)^2}{\mw^4} - \dd\frac{\aww(0)\Vw}{\mw^2} \right\} ~~,
\label{e2.1b}
\eeq
\een
where $e$ and $\mw$ are  the electric charge and the $W$ mass, respectively, 
$s^2 \equiv \sin^2 \theta_\smallw$, $\theta_\smallw$ being the weak 
interaction mixing angle, $\aww$ is the transverse part of the 
$W$ \self, and   $\Vw$ and $\Bw$ represent the relevant
vertex and box corrections. In \equ{e2.1} the subscript $0$ indicates that we 
are dealing with unrenormalized quantities. To express \gmu\ in terms of 
renormalized
parameters,  we insert in \equ{e2.1}\  $e_0^2 = e^2 - \de, \,
s^2_0 = s^2 - \ds, \, \mwb^2 = \mw^2 - \dmw = \mw^2 - \Re \aww (\mw^2)$.  
Our choice of $\dmw$ identifies $\mw$ as the physical mass while,
for the moment, we do not specify the renormalized parameters $e^2$
and $s^2$, but assume that the counterterms \de, \ds\  can contain mass
singularity and  $\mt$-power corrections. After some simple algebra,
and using the fact that at one-loop
$(\Re \aww (\mw^2) - \aww(0))/ \mw^2$ and $\Vw$ do not contain
mass singularities or $\mt^2$ corrections, \equ{e2.1} becomes
\beq
\frac{\gmu}{\sqrt2}  = \frac{e^2}{8 s^2\,\mw^2} & \left\{ 
1+ \frac{\Re \aww (\mw^2)}{\mw^2} -\frac{\aww(0)}{\mw^2} + \Vw + \mw^2 \Bw -
 \deoe + \dsos \right. \nn \\
& - 2 \left( \deoe - \dsos  \right) \left[ 
 \frac{\Re \aww (\mw^2)}{\mw^2} -\frac{\aww(0)}{\mw^2} + \Vw + \mw^2 \Bw
\right] \nn \\
& \left.+ \left( \deoe \right)^2  + 2 \left( \dsos \right)^2 
- 2 \deoe \dsos  \right\},
\label{e2.2}
\eeq
with the understanding that the one-loop contribution is now written in terms
of the renormalized parameters $e^2$ and $s^2$. In \equ{e2.2}, the second
and third lines take into account explicitly the expansion of the overall
coupling $e_0^2/s_0^2$ in the lowest order contribution, while  mass
counterterm effects and shifts in additional $s_0^2$
are included by definition in the two-loop terms. From \equ{e2.2}
it is easy to see  that the replacement in \equ{e2.1a} \cite{Si84}
\be
 1 + \dr \rightarrow \frac{1}{1- \dr} \label{e2.3}
\ee
 takes correctly into account  the  $ \ln (\mz/m_f)$ 
terms contained in $\delta e^2/e^2$, to $O(\alpha^2 \ln (\mz/m_f)$, 
once the renormalized parameter $e$ is identified with the electric charge at
zero momentum transfer. However, as emphasized by Consoli, Hollik,
 and Jegerlehner \cite{CHJ}, there is a mismatch in the iteration 
of the one-loop $\delta s^2 / s^2$ term. This contribution in the OS scheme 
contains  finite
corrections proportional to $\mt^2$ and therefore the replacement
 (\ref{e2.3}) 
does  not take into account correctly  the reducible contribution of 
$O(\alpha^2 \mt^4/\mw^4)$  and 
$O(\alpha^2 \mt^2/\mw^2)$. A way to circumvent  this problem is
to use  an $\msbar$ subtraction
for the parameter $s$, namely to choose the counterterm \ds\ to 
subtract just the
terms proportional to $\delta = (n-4)^{-1} + [\gamma-\ln(4\pi)]/2$
\cite{DFS}. As
 \ds\ does not contain any finite part, this procedure automatically
takes into account all reducible contributions.

The above discussion tells us that the simplest way to take into account the
$O(\alpha^2 \mt^2/\mw^2)$ corrections to the relation between the $\mu$-decay
constant and the W mass is through an $\msbar$ subtraction for the weak
interaction angle. The relations between the $\msbar$ and the OS frameworks 
were worked out in \efe{DFS}. Here we just recall the basic corrections
of the $\msbar$ framework that enter into the $\mw$--$\mz$ interdependence.
They are $\drcar_W$, that relates $G_\mu$ to the $\msbar$
weak interaction angle defined at the scale $\mz$, \sincur\ \cite{Si89},
henceforth abbreviated as \scur, with $\ccur \equiv 1 - \scur$, 
\beq
\frac{G_\mu}{\sqrt{2}}= \frac{\pi \alpha}{2 \scur \mw^2}\frac1{1-\drcar_W},
\label{e2.4}
\eeq
and  $\hat{\rho}$ 
\ben \label{e2.5}
\beq
\hat{\rho}\, = \, \frac{\mw^2}{\mz^2\, \ccur}\, \equiv\, \frac{c^2}{\ccur},
 \label{e2.5a}
\eeq
that is given  explicitly by
\beq
\hat{\rho}=\frac1{1-\Delta\hat{\rho}}=\frac1{1-Y_{\msbar}} \label{e2.5b}
\eeq
with
\beq
        Y_{\msbar} = \frac{1}{\hat{\rho}\, \mz^2}\Re\left[
         \frac{\aww(\mw^2)}{\ccur} - \azz(\mz^2) \right]_{\msbar} .
\label{e2.5c}
\eeq
\een
In \equ{e2.5c} $\azz (\mz^2)$ is the transverse $Z$ \self\ evaluated at the
physical $Z$ mass, the subscript $\msbar$ indicates both the  $\msbar $ 
subtraction and the choice $\mu = \mz$ for the 't Hooft mass scale, and 
we have neglected small contributions proportional to the $\gamma\,Z$
mixing in the $Z$ mass counterterm that do not contain mass singularities
or terms proportional to $\mt^2$. In terms of these
corrections, the  $\mw$--$\mz$ interdependence can be expressed as
\be
        \frac{\mw^2}{\mz^2} = \frac{\hat{\rho}}{2} \left\{
      1 + \left[ 1 - \frac{4 A^2}{\mz^2\hat{\rho}(1-\drcar_W)} \right]^{1/2}
                            \right\} . \label{e2.6}
\ee
where $A =
\left(\pi\alpha/(\sqrt{2}\gmu)\right)^{1/2} = (37.2803\pm 0.0003)$ GeV.

We now start discussing the \amtd\ corrections to $\drcar_W$. Comparing
\equ{e2.2}\ and \equ{e2.4}, and keeping in mind the replacement (\ref{e2.3}),
we see that contributions of
this order come not only from the W and photon two-point functions
(the latter is included in 
$\delta e^2/e^2$), but also  from vertex and box diagrams.
As explained in \efe{us}, by writing the one-loop result in terms of 
$\msbar$ coupling and physical $W$ and $Z$ masses, one automatically 
takes into
account the $\amtd$ corrections coming from the box diagrams and, to a
large extent, the similar contribution coming from the vertices.  
Only the vertex diagrams involving
 a mixing between
vector bosons and unphysical scalars through a fermionic blob have to be 
explicitly calculated.
They  can be easily expressed, however, in terms of
 two-loop self-energy integrals at zero momentum transfer, and 
therefore computed on the same footing as the self-energy contribution.


Because of the presence of the two-loop $W$
mass counterterm, the calculation of $\drcar_W$ involves the evaluation of 
two-loop self-energy integrals both at $q^2 =0$, 
and at $q^2 = \mw^2$, where $q$ is the momentum transfer. Similarly, 
 \eqs{e2.5} show that $\hat{\rho}$ entails 
 two-point functions evaluated at non-zero
momentum transfer.
 Two-loop self-energy diagrams with non-vanishing
masses and  momenta  cannot  
in general be expressed in terms of known functions like polylogarithms.
However, the extraction of the leading $\mt^4$ and next-to-leading
$\mt^2$ contributions from a
two-loop self-energy diagram at non-zero $q^2$  can be performed through 
an asymptotic expansion of the corresponding integrals
in inverse powers of the top mass  \cite{expa}. 
The $q^2 =0$ self-energy integrals, instead, 
 can be exactly solved for any mass,
 expressed in a  closed form \cite{DT}, and then  
expanded in top mass powers. The  zero momentum transfer contributions
of the $W$ and $Z$ self-energies have been explicitly checked 
with   the Ward identities of the theory, 
which were
derived up to \amtd\ by current algebra methods \cite{DFG,thesis}. In the
calculation we consider the Higgs mass as a free parameter. Therefore,
before performing the heavy top expansion, we need to specify whether
$\mh$ can be considered light with respect to $\mt$, or $\mh,\, \mt \gg
\mz$, with an arbitrary ratio $\mh / \mt$. Consequently, we derive 
expressions for the two-loop corrections in the two regions.

The identification of the $\mt^2$ two-loop contribution to $\drcar_W$
and  $\Delta\hat{\rho}$ requires a precise specification of the one-loop 
term. 
Our one-loop contribution  coincides with the
analytic expressions reported in \efe{DFS}, 
written in terms of physical masses  and  couplings $\alpha$ and $\ccur$
($\drcar_W$), and $\acur$, the $\msbar$ e.m.~running coupling, 
and $\ccur$ ($\Delta\hat{\rho}$).

With this convention for the one-loop term, we find for the two-loop \amtd\
contribution to $\drcar_W$, in units 
$ N_c \,(\alpha/(4 \pi \hat{s}^2))^2 \, \, \mt^2/\mw^2 $, with 
the colour factor $N_c=3$, 
\ben\label{drcw}
\beq
\Delta\hat{r}_W^{(2)} &=
-{{13}\over {144}} - {1\over {48{\chat^4}}} - {{41}\over {96\,{\ccur}}} +
   \frac{61 \ccur}{72}+ 
  \frac{7-16 \ccur}{27} \pi\,\sqrt{\it ht}  - {{{{\pi }^2}}\over {36}} - 
  {{5\,{{{\it ht}}^2}}\over {144\,{\chat^4}\,{{{\it zt}}^2}}} 
+   {{35\,{\it ht}}\over {288\,{\ccur}\,{\it zt}}} \nn\\&+ 
   {5\over 12} \left(1 + 
{{{{{\it ht}}^2}}\over {12\,{\chat^4}\,{{\it zt}^2}}} 
- {{{\it ht}}\over {3\,{\ccur}\,{\it zt}}} \right) 
   B0[\mw^2,\mh^2,\mw^2] + 
\frac{1+20\chat^2-24 \chat^4}{48\chat^4}
    B0[\mw^2,\mw^2,\mz^2] \nn\\&
-   {{\left( 5 \hat{s}^2 {{{\it ht}}^2}  + 
        3\,{\it ht}\,{\it zt} + 48\,{\ccur}\,{\it ht}\,{\it zt} - 
        60\,{\chat^4}\,{\it ht}\,{\it zt} - 3\,{\ccur}\,{{{\it zt}}^2} - 
        8\,{\chat^4}\,{{{\it zt}}^2} + 20\,{\chat^6}\,{{{\it zt}}^2} 
\right) \,
      \ln {\ccur}}\over 
    {144\,{\ccur}\, \scur \,{\it zt}\,
      \left( {\it ht} - {\ccur}\,{\it zt} \right) }} \nn\\&+ 
  {{5\,{\it ht}\,\left( {{{\it ht}}^2} - 4\,{\ccur}\,{\it ht}\,{\it zt} + 
        12\,{\chat^4}\,{{{\it zt}}^2} \right) \,\ln {\it ht}}\over 
    {144\,{\chat^4}\,{{{\it zt}}^2}\,\left( {\it ht} - {\ccur}\,{\it zt}
 \right) }} 
   + \left( {17\over 36} 
  - \frac{13\ccur}{18} \right) 
   \ln {\mt^2\over \mu^2} \nn\\&- 
  {{\left( 5\,{\ccur}\,{{{\it ht}}^2} - 3\,{\it ht}\,{\it zt} - 
        60\,{\ccur}\,{\it ht}\,{\it zt} + 60\,{\chat^4}\,{\it ht}\,{\it zt} + 
        (3\,{\ccur} + 60\,{\chat^4} - 
        20\,{\chat^6})\,{{{\it zt}}^2} \right) \ln {\it zt}}\over 
   {144\,{\chat^4}\,{\it zt}\,\left( {\it ht} - {\ccur}\,{\it zt} \right) }},
\eeq
for a light Higgs expansion, $\mh \ll \mt$, while in the region $\mh \gg \mz$
we obtain 
\beq\label{drwlh}
\Delta\hat{r}_W^{(2)} &=
-{{121}\over {288}} - {1\over {48\,{\chat^4}}} - {{41}\over {96\,{\ccur}}} + 
  \frac{77\,\ccur}{12}+{{19}\over {72\,{\it ht}}} + \left(\frac{41}{216} - 
  \frac{4 \,\ccur}{27}\right){\it ht} - 
  {{\left( 19 + 21\,{\it ht} \right) \,{{\pi }^2}}\over 
    {432\,{{{\it ht}}^2}}} \nn\\&
- \left( {1\over 2} - {1\over {48\,{\chat^4}}} - 
     {5\over {12\,{\ccur}}} \right)   B0[\mw^2,\mw^2,\mz^2] +
 \frac{16\, \ccur-7}{216} (ht-4)   
  \sqrt{\it ht}\, g({\it ht}) \nn\\&
-  \left( {1\over {12}} - {1\over {3\,{\it ht}}} \right) \,
    \Lambda({\it ht}) 
 + {\left( 19 + 21 ht - 12 ht^2 - 
        31\,ht^3 + 9 \, ht^4 \right) 
\over {72\,ht^2}}      {\rm Li_2}(1 - {\it ht}) \nn\\&- 
  {{\left( 1 + 21\,{\ccur} - 25\,{\chat^4} \right) \,\ln {\ccur}}\over 
    {48\,{\ccur}\, \scur}} 
+  \left( {17\over 36}- \frac{13\,\ccur}{18} \right) \,
   \ln {\mt^2\over \mu^2} + 
{  {\left(1 + 20\,{\ccur} - 25\,{\chat^4} \right)\,\ln {\it zt}}\over 
    {48\,{\chat^4}}}
\nn\\&
+ \frac{372+(96\ccur-213) ht+ (432 \ccur-318 )ht^2 + 
(97 - 160\ccur) ht^3 -(7-16\ccur)
  ht^4  }{216(ht-4)ht} \,\ln {\it ht} 
 \nn\\&+ 
\frac{ 96 -(384-64 \ccur){\it ht} -(2+64 \ccur) ht^2 + 
        231\,{{{\it ht}}^3} -85 ht^4+ 9 ht^5}
 {144\,(ht-4) ht^2} \,\phi(\frac{\it ht}{4}).
\eeq
\een
In \eqs{drcw} $ht \equiv (\mh/\mt)^2$, $zt \equiv (\mz/\mt)^2$,
\ben \label{e2.15} \beq
g(x) =&  \left\{
          \begin{array}{lr}
           \sqrt{4-x}\,\left
              (\pi - 2 \arcsin{\sqrt{x/4}}
                       \right) &                        0 < x \leq 4 \\
           {}\\
           2 \sqrt{x/4-1}\,\ln\left(
                \frac{1-\sqrt{1-4/x}}{1+\sqrt{1-4/x}}
                              \right) &                 x > 4 \,~ ,
          \end{array}
        \right.  \label{e2.15a}
\eeq
\beq
\Lambda( x) =&  \left\{
          \begin{array}{lr}
          -\frac{1}{2 \sqrt{x}}\,g(x) + {\pi \over 2} \,\sqrt{4/x -1}
                      &                         0 < x \leq 4 \\
           {}\\
            -\frac{1}{2 \sqrt{x}}\, g(x)&               x > 4 \,~ ,
          \end{array}
        \right.  \label{e2.15abis}
\eeq
we have indicated the dilogarithmic function as 
${\rm{Li_2}} (x) = - \int_0^x dt \, {\ln (1-t) \over t} $, 
and introduced 
\be
\phi(z) =
       \begin{cases}
       4 \sqrt{{z \over 1-z}} ~Cl_2 ( 2 \arcsin \sqrt z ) & $ 0 < z
\leq 1$\\
       { 1 \over \lambda} \left[ - 4 {\rm Li_2} ({1-\lambda \over 2}) +
       2 \ln^2 ({1-\lambda \over 2}) - \ln^2 (4z) +\pi^2/3 \right] 
       & $z >1 $\,,
       \end{cases}
       \label{e2.15c}
\ee \een
where $Cl_2(x)= {\rm Im} \,{\rm Li_2} (e^{ix})$ is the Clausen function
with
$\lambda = \sqrt{1 - {1 \over z}}$.
The function \\ $B0[q^2,m_1^2,m_2^2]$ is defined through the one-loop 
integral ($\eps= (4-n)/2$)
\beq
-i \,\mt^{2\epsilon} e^{\gamma\epsilon}\int \frac{d^n k}{\pi^{n/2}}
\frac1{[k^2-m_1^2][(k-q)^2-m_2^2]}=
\frac1{\epsilon} + B0[q^2,m_1^2,m_2^2] + O(\epsilon),
\label{B0}
\eeq
whose analytic form is well known (see for example \efe{DS}).
 It is interesting 
to note that the $O(\eps)$ part of one-loop integrals like the one in 
\eqs{B0} cancel exactly in the final two-loop expression.

The two-loop contribution to $\drcar_W$ is quite small over the whole range 
of realistic top and Higgs mass values. For instance, using 
$\mt=180$ GeV and 
$\scur=0.2314$, we find that $\drcar_W^{(2)}$ has  an  absolute maximum
at $\mh=0$, $+5.7\times 10^{-5}$, then decreases very rapidly  
for increasing  $\mh$.
The two expansions \eqs{drcw} meet at $\mh\simeq 0.3\, \mt$,
and for the whole range 65GeV$<\mh <$1TeV, 
$|\drcar_W^{(2)}|<1\times 10^{-5}$.
 The same happens for different but realistic
values of $\mt$. This is indeed a quite small effect, comparable in size to 
routinely neglected contributions.

The calculation of  $\Delta\hat{\rho}^{(2)}$
 yields, in units $N_c\, [\acur\,\mt^2/(16 \pi \scur \ccur\mz^2)
]^2 \simeq N_c x_t^2$ ($x_t= \\ G_\mu \mt^2/(8 \pi^2 \sqrt{2})$):
\ben \label{e2.14} \beq
\Delta\hat{\rho}^{(2)}&=
19 - {{53\,{\it ht}}\over 3} + {{3\,{{{\it ht}}^{{3\over 2}}}\,\pi }\over 
    2} + {{8\,{{\it ht}^2}}\over {9\,{\it zt}}} - 
  {{5\,{{{\it ht}}^2}}\over {9\,{\ccur}\,{\it zt}}}
 +   \left({{845\over {27}}} - {1\over {3\,{\ccur}}} + 
  {{427\,{\ccur}\over {27}}} - {{122\,{\chat^4}\over 9}}\right)zt
\nn\\& + 
  \frac{\pi^2}{27}\left(  54\,{\it ht}-54 - 119\,{\it zt} + 
        44\,{\ccur}\,{\it zt} \right)  +
  \frac4{27}\,{{{\sqrt{{\it ht}}}\,\pi 
      \left( -27 + 34\,{\it zt} - 116\,{\ccur}\,{\it zt} + 
        64\,{\chat^4}\,{\it zt} \right) }}\nn\\& + 
  \left( {{32{\it ht}}\over 9} - {{8{{{\it ht}}^2}}\over {9{\it zt}}} - 
     {{32{\it zt}}\over 3} \right) 
    B0[\mz^2,\mh^2,\mz^2]
+   {{\left( 1 + 20{\ccur} - 24{\chat^4} \right) {{\it zt}\over 3 \ccur}}}
      B0[\mw^2,\mw^2,\mz^2]
 \nn\\&
- {2\over3} ( 1 + 18 \ccur - 16{\chat^4} ) {\it zt}
      B0[\mz^2,\mw^2,\mw^2] - \frac5{9} \left( 4{\it ht} - 
     {{\it ht}^2\over {{\ccur}{\it zt}}} -
      12 \ccur {\it zt} \right) 
   B0[\mw^2,\mh^2,\mw^2] \nn\\&
- \frac1{9}\,\left( 5\,{\it ht} + 3\,{\it zt} + 32\,{\ccur}\,{\it zt} + 
        48\,{\chat^4}\,{\it zt} \right) \,\ln {\ccur} + 
  {{\frac{\it ht}{9 \ccur {\it zt}}\,\left( 5\,{\it ht} - 8\,{\ccur}
\,{\it ht} - 
        18\,{\ccur}\,{\it zt} \right) \ln {\it ht}}} \nn\\&
+\frac8{9} \left( 4 - 26\,{\ccur} - 5\,{\chat^4} \right) zt 
\, \ln{\mt^2\over \mu^2} 
+   \left( {{\it ht}\over 3} - {{11\,{\it zt}}\over 9} + 
     {{{\it zt}}\over {3\,{\ccur}}} - {{16\,{\ccur}\,{\it zt}}\over 9} - 
     {{16\,{\chat^4}\,{\it zt}}\over 3} \right) \ln {\it zt},
\label{e2.14a}
\eeq
for a light Higgs $\mh\ll \mt$. A heavy Higgs expansion gives instead
\beq
\Delta \hat{\rho}^{(2)} &= 
25 - 4{\it ht} + \left( {1\over 2} - {1\over {{\it ht}}} \right) 
   {{\pi }^2} + {{\left( {\it ht}-4 \right) {\sqrt{{\it ht}}}\,
       g(\it ht)}\over 2} 
+ \left( -6 - 6\,{\it ht} + 
     {{{{{\it ht}}^2}}\over 2} \right) \,\ln {\it ht} \nn\\&
+  \left(  {6\over {{\it ht}}} -15+ 12 {\it ht} - 3\,{{{\it ht}}^2} \right)
{\rm Li_2}(1 - {\it ht}) +
  \frac3{2} ( -10 + 6\,{\it ht} - {{{\it ht}}^2} ) 
\,      \phi ({{\it ht}\over 4}) \nn\\&+ 
  {\it zt}\,\left[  \frac1{54 {\hat{c}^2}\,( {\it ht}-4) \,{\it ht}}
   \left( -1776\,{\hat{c}^4} + (72 - 6250\,{\hat{c}^2} - 
         3056\,{\hat{c}^4} + 3696{\hat{c}^6})\,{\it ht}
\right. \right.\nn \\&
\left.  +(- 18 + 
         1283\,{\hat{c}^2} + 1371{\hat{c}^4}
 -   1436\,{\hat{c}^6})\,{{\it ht}^2} + (68\,{\hat{c}^2} - 
      124{\hat{c}^4} + 128{\hat{c}^6})\,{{{\it ht}}^3}\right)\nn\\&
 + 
     {{\left(6{\hat{c}^2}{\it ht} -37{\hat{c}^2}- 119\,{{{\it ht}}^2} + 
           56\,{\hat{c}^2}\,{{{\it ht}}^2} \right) {{\pi }^2}}\over 
       {27\,{{{\it ht}}^2}}} + \left(    {{32\,{\hat{c}^4}}\over 3}
-{2\over 3} - 12\,{\hat{c}^2}       \right) 
      B0[\mz^2,\mw^2,\mw^2] \nn\\&+ 
    \left( {{20}\over 3} + {1\over {3\,{\hat{c}^2}}} - 8\,{\hat{c}^2} \right)
       B0[\mw^2,\mw^2,\mz^2] + 
     {{\left( 17 - 58\,{\hat{c}^2} + 32\,{\hat{c}^4} \right) 
         \left( 4 - {\it ht} \right) {\sqrt{{\it ht}}}\,
   g(\it ht)}\over {27}} \nn\\&
- {{40\scur 
         \left( 4 - {\it ht} \right) { \Lambda}({\it ht})}\over 
       {3\,{\it ht}}} 
+ {{2\,{\hat{c}^2} \left( 37 - 6\,{\it ht} - 
12\,{{{\it ht}}^2} - 22\,{{{\it ht}}^3} + 
           9\,{{{\it ht}}^4} \right) \,{\rm Li_2}(1 - {\it ht})}\over 
       {9{{{\it ht}}^2}}} \nn\\&
- {{\left( 1 + 14\,{\hat{c}^2} + 16\,{\hat{c}^4} \right) 
         \ln {\hat{c}^2}}\over 3}
 +   \left( 11520 - 15072\,{\hat{c}^2} -( 7170
 - 8928{\hat{c}^2} - 
           768{\hat{c}^4}){\it ht} \right.\nn\\&
+(3411 -    7062{\hat{c}^2} + 3264\,{\hat{c}^4}){{{\it ht}}^2} -( 
           1259 - 3547\,{\hat{c}^2}+ 
           2144{\hat{c}^4}){{{\it ht}}^3} \nn\\&
+ (238 -            758{\hat{c}^2} + 448\,{\hat{c}^4}){{{\it ht}}^4}
 -\left.(            17 - 58\,{\hat{c}^2} + 
           32{\hat{c}^4}){{{\it ht}}^5} \right) {{\ln {\it ht}}\over 
       {27{{\left( {\it ht} -4\right) }^2}\,{\it ht}}} \nn\\&
+      \frac8{9} \left( 4 - 26\,{\ccur} - 5\,{\chat^4} \right)
         \ln {\mt^2\over\mu^2} 
+     {{\left( 3 + 5\,{\hat{c}^2} - 26\,{\hat{c}^4} -
 48\,{\hat{c}^6} \right)   \ln {\it zt}}\over {9\,{\hat{c}^2}}} 
\nn\\&+ 
     \left( 3840 \scur -( 4310 - 4224\,{\hat{c}^2} - 
           256{\hat{c}^4})\,{\it ht} + (1706 - 
          1312\,{\hat{c}^2} - 320\,{\hat{c}^4}) ht^2 \right.\label{e2.14b}\\&
\left.\left.-( 315 + 476\,{\hat{c}^2} - 
           64{\hat{c}^4})ht^3+ (24 + 
        454\,{\hat{c}^2})\,{{{\it ht}}^4} - 112\,{\hat{c}^2}\,{{\it ht}^5} +
         9\,{\hat{c}^2}\,{{{\it ht}}^6} \right) 
{{\phi ({{\it ht}\over 4})}\over
         {9{{\left( {\it ht}-4 \right) }^2}{{{\it ht}}^2}}}
 \right]. \nn
\eeq
\een

The first two lines of \eqs{e2.14b} represent the leading $\gmuq$
result \cite{bar,DFG}, which is completely independent of the gauge
sector of the theory. Indeed this part can be computed in the
framework of a pure Yukawa theory, obtained from the SM in the limit
of vanishing gauge coupling constants.
The rest of \equ{e2.14a} is proportional to $zt=\mz^2/\mt^2$
and represents the first correction to the Yukawa limit.

We note that the \amtd\ contribution to \eqs{e2.14} is much more relevant 
than the one to $\drcar_W$, following a pattern 
very similar to the one for the $\rho$ parameter in $\nu_\mu-e$ scattering
\cite{DFG,us}.
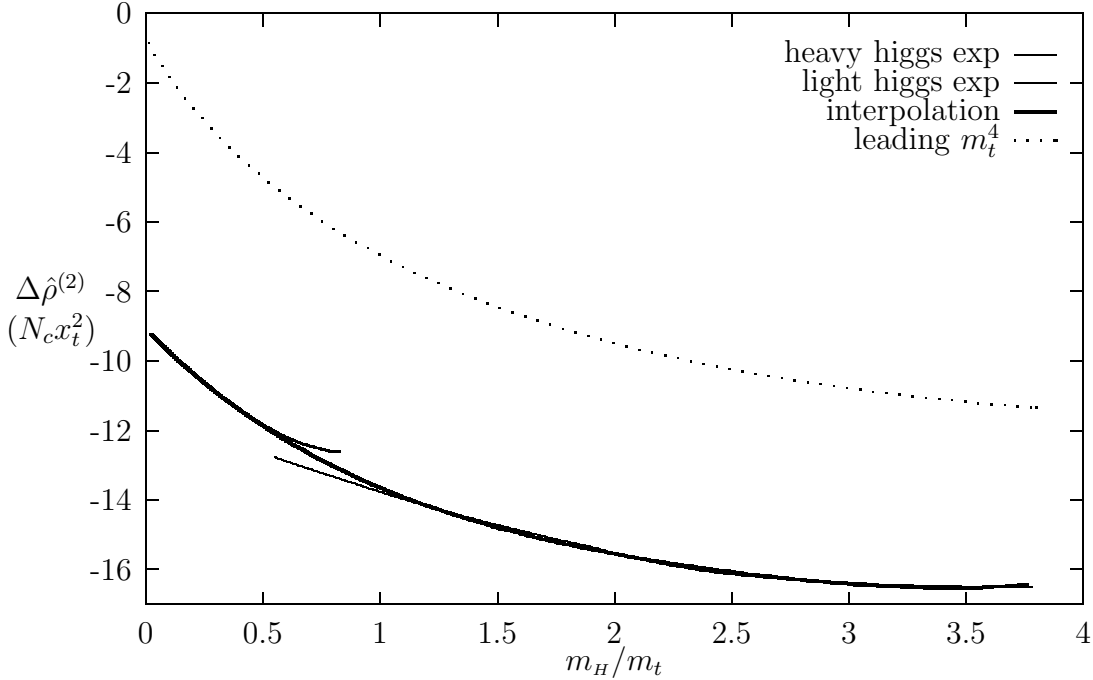
\begin{figure}   
\setlength{\unitlength}{0.240900pt}
\ifx\plotpoint\undefined\newsavebox{\plotpoint}\fi
\sbox{\plotpoint}{\rule[-0.175pt]{0.350pt}{0.350pt}}%
\begin{picture}(1800,1200)(0,0)
\sbox{\plotpoint}{\rule[-0.175pt]{0.350pt}{0.350pt}}%
\put(264,158){\rule[-0.175pt]{0.350pt}{223.796pt}}
\put(264,213){\rule[-0.175pt]{4.818pt}{0.350pt}}
\put(242,213){\makebox(0,0)[r]{-16}}
\put(1716,213){\rule[-0.175pt]{4.818pt}{0.350pt}}
\put(264,322){\rule[-0.175pt]{4.818pt}{0.350pt}}
\put(242,322){\makebox(0,0)[r]{-14}}
\put(1716,322){\rule[-0.175pt]{4.818pt}{0.350pt}}
\put(264,431){\rule[-0.175pt]{4.818pt}{0.350pt}}
\put(242,431){\makebox(0,0)[r]{-12}}
\put(1716,431){\rule[-0.175pt]{4.818pt}{0.350pt}}
\put(264,541){\rule[-0.175pt]{4.818pt}{0.350pt}}
\put(242,541){\makebox(0,0)[r]{-10}}
\put(1716,541){\rule[-0.175pt]{4.818pt}{0.350pt}}
\put(264,650){\rule[-0.175pt]{4.818pt}{0.350pt}}
\put(242,650){\makebox(0,0)[r]{-8}}
\put(1716,650){\rule[-0.175pt]{4.818pt}{0.350pt}}
\put(264,759){\rule[-0.175pt]{4.818pt}{0.350pt}}
\put(242,759){\makebox(0,0)[r]{-6}}
\put(1716,759){\rule[-0.175pt]{4.818pt}{0.350pt}}
\put(264,868){\rule[-0.175pt]{4.818pt}{0.350pt}}
\put(242,868){\makebox(0,0)[r]{-4}}
\put(1716,868){\rule[-0.175pt]{4.818pt}{0.350pt}}
\put(264,978){\rule[-0.175pt]{4.818pt}{0.350pt}}
\put(242,978){\makebox(0,0)[r]{-2}}
\put(1716,978){\rule[-0.175pt]{4.818pt}{0.350pt}}
\put(264,1087){\rule[-0.175pt]{4.818pt}{0.350pt}}
\put(242,1087){\makebox(0,0)[r]{0}}
\put(1716,1087){\rule[-0.175pt]{4.818pt}{0.350pt}}
\put(264,158){\rule[-0.175pt]{0.350pt}{4.818pt}}
\put(264,113){\makebox(0,0){0}}
\put(264,1067){\rule[-0.175pt]{0.350pt}{4.818pt}}
\put(448,158){\rule[-0.175pt]{0.350pt}{4.818pt}}
\put(448,113){\makebox(0,0){0.5}}
\put(448,1067){\rule[-0.175pt]{0.350pt}{4.818pt}}
\put(632,158){\rule[-0.175pt]{0.350pt}{4.818pt}}
\put(632,113){\makebox(0,0){1}}
\put(632,1067){\rule[-0.175pt]{0.350pt}{4.818pt}}
\put(816,158){\rule[-0.175pt]{0.350pt}{4.818pt}}
\put(816,113){\makebox(0,0){1.5}}
\put(816,1067){\rule[-0.175pt]{0.350pt}{4.818pt}}
\put(1000,158){\rule[-0.175pt]{0.350pt}{4.818pt}}
\put(1000,113){\makebox(0,0){2}}
\put(1000,1067){\rule[-0.175pt]{0.350pt}{4.818pt}}
\put(1184,158){\rule[-0.175pt]{0.350pt}{4.818pt}}
\put(1184,113){\makebox(0,0){2.5}}
\put(1184,1067){\rule[-0.175pt]{0.350pt}{4.818pt}}
\put(1368,158){\rule[-0.175pt]{0.350pt}{4.818pt}}
\put(1368,113){\makebox(0,0){3}}
\put(1368,1067){\rule[-0.175pt]{0.350pt}{4.818pt}}
\put(1552,158){\rule[-0.175pt]{0.350pt}{4.818pt}}
\put(1552,113){\makebox(0,0){3.5}}
\put(1552,1067){\rule[-0.175pt]{0.350pt}{4.818pt}}
\put(1736,158){\rule[-0.175pt]{0.350pt}{4.818pt}}
\put(1736,113){\makebox(0,0){4}}
\put(1736,1067){\rule[-0.175pt]{0.350pt}{4.818pt}}
\put(264,158){\rule[-0.175pt]{354.605pt}{0.350pt}}
\put(1736,158){\rule[-0.175pt]{0.350pt}{223.796pt}}
\put(264,1087){\rule[-0.175pt]{354.605pt}{0.350pt}}
\put(45,622){\makebox(0,0)[l]{\shortstack{$\Delta\hat{\rho}^{(2)}$\\
 $(N_c x_t^2)$}}}
\put(1000,68){\makebox(0,0){$\mh/\mt$}}
\put(264,158){\rule[-0.175pt]{0.350pt}{223.796pt}}
\put(1606,1022){\makebox(0,0)[r]{heavy higgs exp}}
\put(1628,1022){\rule[-0.175pt]{15.899pt}{0.350pt}}
\put(466,389){\usebox{\plotpoint}}
\put(466,389){\rule[-0.175pt]{0.704pt}{0.350pt}}
\put(468,388){\rule[-0.175pt]{0.704pt}{0.350pt}}
\put(471,387){\rule[-0.175pt]{0.704pt}{0.350pt}}
\put(474,386){\rule[-0.175pt]{0.704pt}{0.350pt}}
\put(477,385){\rule[-0.175pt]{0.704pt}{0.350pt}}
\put(480,384){\rule[-0.175pt]{0.704pt}{0.350pt}}
\put(483,383){\rule[-0.175pt]{0.704pt}{0.350pt}}
\put(486,382){\rule[-0.175pt]{0.704pt}{0.350pt}}
\put(489,381){\rule[-0.175pt]{0.704pt}{0.350pt}}
\put(492,380){\rule[-0.175pt]{0.704pt}{0.350pt}}
\put(495,379){\rule[-0.175pt]{0.704pt}{0.350pt}}
\put(498,378){\rule[-0.175pt]{0.704pt}{0.350pt}}
\put(501,377){\rule[-0.175pt]{0.704pt}{0.350pt}}
\put(503,376){\rule[-0.175pt]{0.743pt}{0.350pt}}
\put(507,375){\rule[-0.175pt]{0.743pt}{0.350pt}}
\put(510,374){\rule[-0.175pt]{0.743pt}{0.350pt}}
\put(513,373){\rule[-0.175pt]{0.743pt}{0.350pt}}
\put(516,372){\rule[-0.175pt]{0.743pt}{0.350pt}}
\put(519,371){\rule[-0.175pt]{0.743pt}{0.350pt}}
\put(522,370){\rule[-0.175pt]{0.743pt}{0.350pt}}
\put(525,369){\rule[-0.175pt]{0.743pt}{0.350pt}}
\put(528,368){\rule[-0.175pt]{0.743pt}{0.350pt}}
\put(531,367){\rule[-0.175pt]{0.743pt}{0.350pt}}
\put(534,366){\rule[-0.175pt]{0.743pt}{0.350pt}}
\put(537,365){\rule[-0.175pt]{0.743pt}{0.350pt}}
\put(540,364){\rule[-0.175pt]{0.743pt}{0.350pt}}
\put(544,363){\rule[-0.175pt]{0.743pt}{0.350pt}}
\put(547,362){\rule[-0.175pt]{0.743pt}{0.350pt}}
\put(550,361){\rule[-0.175pt]{0.743pt}{0.350pt}}
\put(553,360){\rule[-0.175pt]{0.743pt}{0.350pt}}
\put(556,359){\rule[-0.175pt]{0.743pt}{0.350pt}}
\put(559,358){\rule[-0.175pt]{0.743pt}{0.350pt}}
\put(562,357){\rule[-0.175pt]{0.743pt}{0.350pt}}
\put(565,356){\rule[-0.175pt]{0.743pt}{0.350pt}}
\put(568,355){\rule[-0.175pt]{0.743pt}{0.350pt}}
\put(571,354){\rule[-0.175pt]{0.743pt}{0.350pt}}
\put(574,353){\rule[-0.175pt]{0.743pt}{0.350pt}}
\put(577,352){\rule[-0.175pt]{0.743pt}{0.350pt}}
\put(581,351){\rule[-0.175pt]{0.743pt}{0.350pt}}
\put(584,350){\rule[-0.175pt]{0.743pt}{0.350pt}}
\put(587,349){\rule[-0.175pt]{0.743pt}{0.350pt}}
\put(590,348){\rule[-0.175pt]{0.743pt}{0.350pt}}
\put(593,347){\rule[-0.175pt]{0.743pt}{0.350pt}}
\put(596,346){\rule[-0.175pt]{0.743pt}{0.350pt}}
\put(599,345){\rule[-0.175pt]{0.743pt}{0.350pt}}
\put(602,344){\rule[-0.175pt]{0.743pt}{0.350pt}}
\put(605,343){\rule[-0.175pt]{0.743pt}{0.350pt}}
\put(608,342){\rule[-0.175pt]{0.743pt}{0.350pt}}
\put(611,341){\rule[-0.175pt]{0.743pt}{0.350pt}}
\put(614,340){\rule[-0.175pt]{0.743pt}{0.350pt}}
\put(618,339){\rule[-0.175pt]{0.743pt}{0.350pt}}
\put(621,338){\rule[-0.175pt]{0.743pt}{0.350pt}}
\put(624,337){\rule[-0.175pt]{0.743pt}{0.350pt}}
\put(627,336){\rule[-0.175pt]{0.743pt}{0.350pt}}
\put(630,335){\rule[-0.175pt]{0.743pt}{0.350pt}}
\put(633,334){\rule[-0.175pt]{0.743pt}{0.350pt}}
\put(636,333){\rule[-0.175pt]{0.743pt}{0.350pt}}
\put(639,332){\rule[-0.175pt]{0.743pt}{0.350pt}}
\put(642,331){\rule[-0.175pt]{0.743pt}{0.350pt}}
\put(645,330){\rule[-0.175pt]{0.743pt}{0.350pt}}
\put(648,329){\rule[-0.175pt]{0.743pt}{0.350pt}}
\put(651,328){\rule[-0.175pt]{0.810pt}{0.350pt}}
\put(655,327){\rule[-0.175pt]{0.810pt}{0.350pt}}
\put(658,326){\rule[-0.175pt]{0.810pt}{0.350pt}}
\put(662,325){\rule[-0.175pt]{0.810pt}{0.350pt}}
\put(665,324){\rule[-0.175pt]{0.810pt}{0.350pt}}
\put(668,323){\rule[-0.175pt]{0.810pt}{0.350pt}}
\put(672,322){\rule[-0.175pt]{0.810pt}{0.350pt}}
\put(675,321){\rule[-0.175pt]{0.810pt}{0.350pt}}
\put(678,320){\rule[-0.175pt]{0.810pt}{0.350pt}}
\put(682,319){\rule[-0.175pt]{0.810pt}{0.350pt}}
\put(685,318){\rule[-0.175pt]{0.810pt}{0.350pt}}
\put(689,317){\rule[-0.175pt]{0.832pt}{0.350pt}}
\put(692,316){\rule[-0.175pt]{0.832pt}{0.350pt}}
\put(695,315){\rule[-0.175pt]{0.832pt}{0.350pt}}
\put(699,314){\rule[-0.175pt]{0.832pt}{0.350pt}}
\put(702,313){\rule[-0.175pt]{0.832pt}{0.350pt}}
\put(706,312){\rule[-0.175pt]{0.832pt}{0.350pt}}
\put(709,311){\rule[-0.175pt]{0.832pt}{0.350pt}}
\put(713,310){\rule[-0.175pt]{0.832pt}{0.350pt}}
\put(716,309){\rule[-0.175pt]{0.832pt}{0.350pt}}
\put(720,308){\rule[-0.175pt]{0.832pt}{0.350pt}}
\put(723,307){\rule[-0.175pt]{0.832pt}{0.350pt}}
\put(726,306){\rule[-0.175pt]{0.891pt}{0.350pt}}
\put(730,305){\rule[-0.175pt]{0.891pt}{0.350pt}}
\put(734,304){\rule[-0.175pt]{0.891pt}{0.350pt}}
\put(738,303){\rule[-0.175pt]{0.891pt}{0.350pt}}
\put(741,302){\rule[-0.175pt]{0.891pt}{0.350pt}}
\put(745,301){\rule[-0.175pt]{0.891pt}{0.350pt}}
\put(749,300){\rule[-0.175pt]{0.891pt}{0.350pt}}
\put(752,299){\rule[-0.175pt]{0.891pt}{0.350pt}}
\put(756,298){\rule[-0.175pt]{0.891pt}{0.350pt}}
\put(760,297){\rule[-0.175pt]{0.891pt}{0.350pt}}
\put(764,296){\rule[-0.175pt]{0.891pt}{0.350pt}}
\put(767,295){\rule[-0.175pt]{0.891pt}{0.350pt}}
\put(771,294){\rule[-0.175pt]{0.891pt}{0.350pt}}
\put(775,293){\rule[-0.175pt]{0.891pt}{0.350pt}}
\put(778,292){\rule[-0.175pt]{0.891pt}{0.350pt}}
\put(782,291){\rule[-0.175pt]{0.891pt}{0.350pt}}
\put(786,290){\rule[-0.175pt]{0.891pt}{0.350pt}}
\put(789,289){\rule[-0.175pt]{0.891pt}{0.350pt}}
\put(793,288){\rule[-0.175pt]{0.891pt}{0.350pt}}
\put(797,287){\rule[-0.175pt]{0.891pt}{0.350pt}}
\put(801,286){\rule[-0.175pt]{0.990pt}{0.350pt}}
\put(805,285){\rule[-0.175pt]{0.990pt}{0.350pt}}
\put(809,284){\rule[-0.175pt]{0.990pt}{0.350pt}}
\put(813,283){\rule[-0.175pt]{0.990pt}{0.350pt}}
\put(817,282){\rule[-0.175pt]{0.990pt}{0.350pt}}
\put(821,281){\rule[-0.175pt]{0.990pt}{0.350pt}}
\put(825,280){\rule[-0.175pt]{0.990pt}{0.350pt}}
\put(829,279){\rule[-0.175pt]{0.990pt}{0.350pt}}
\put(833,278){\rule[-0.175pt]{0.990pt}{0.350pt}}
\put(837,277){\rule[-0.175pt]{0.990pt}{0.350pt}}
\put(842,276){\rule[-0.175pt]{0.990pt}{0.350pt}}
\put(846,275){\rule[-0.175pt]{0.990pt}{0.350pt}}
\put(850,274){\rule[-0.175pt]{0.990pt}{0.350pt}}
\put(854,273){\rule[-0.175pt]{0.990pt}{0.350pt}}
\put(858,272){\rule[-0.175pt]{0.990pt}{0.350pt}}
\put(862,271){\rule[-0.175pt]{0.990pt}{0.350pt}}
\put(866,270){\rule[-0.175pt]{0.990pt}{0.350pt}}
\put(870,269){\rule[-0.175pt]{0.990pt}{0.350pt}}
\put(874,268){\rule[-0.175pt]{0.990pt}{0.350pt}}
\put(879,267){\rule[-0.175pt]{0.990pt}{0.350pt}}
\put(883,266){\rule[-0.175pt]{0.990pt}{0.350pt}}
\put(887,265){\rule[-0.175pt]{0.990pt}{0.350pt}}
\put(891,264){\rule[-0.175pt]{0.990pt}{0.350pt}}
\put(895,263){\rule[-0.175pt]{0.990pt}{0.350pt}}
\put(899,262){\rule[-0.175pt]{0.990pt}{0.350pt}}
\put(903,261){\rule[-0.175pt]{0.990pt}{0.350pt}}
\put(907,260){\rule[-0.175pt]{0.990pt}{0.350pt}}
\put(911,259){\rule[-0.175pt]{1.144pt}{0.350pt}}
\put(916,258){\rule[-0.175pt]{1.144pt}{0.350pt}}
\put(921,257){\rule[-0.175pt]{1.144pt}{0.350pt}}
\put(926,256){\rule[-0.175pt]{1.144pt}{0.350pt}}
\put(931,255){\rule[-0.175pt]{1.144pt}{0.350pt}}
\put(935,254){\rule[-0.175pt]{1.144pt}{0.350pt}}
\put(940,253){\rule[-0.175pt]{1.144pt}{0.350pt}}
\put(945,252){\rule[-0.175pt]{1.144pt}{0.350pt}}
\put(950,251){\rule[-0.175pt]{0.990pt}{0.350pt}}
\put(954,250){\rule[-0.175pt]{0.990pt}{0.350pt}}
\put(958,249){\rule[-0.175pt]{0.990pt}{0.350pt}}
\put(962,248){\rule[-0.175pt]{0.990pt}{0.350pt}}
\put(966,247){\rule[-0.175pt]{0.990pt}{0.350pt}}
\put(970,246){\rule[-0.175pt]{0.990pt}{0.350pt}}
\put(974,245){\rule[-0.175pt]{0.990pt}{0.350pt}}
\put(978,244){\rule[-0.175pt]{0.990pt}{0.350pt}}
\put(982,243){\rule[-0.175pt]{0.990pt}{0.350pt}}
\put(986,242){\rule[-0.175pt]{0.891pt}{0.350pt}}
\put(990,241){\rule[-0.175pt]{0.891pt}{0.350pt}}
\put(994,240){\rule[-0.175pt]{0.891pt}{0.350pt}}
\put(998,239){\rule[-0.175pt]{0.891pt}{0.350pt}}
\put(1001,238){\rule[-0.175pt]{0.891pt}{0.350pt}}
\put(1005,237){\rule[-0.175pt]{0.891pt}{0.350pt}}
\put(1009,236){\rule[-0.175pt]{0.891pt}{0.350pt}}
\put(1012,235){\rule[-0.175pt]{0.891pt}{0.350pt}}
\put(1016,234){\rule[-0.175pt]{0.891pt}{0.350pt}}
\put(1020,233){\rule[-0.175pt]{0.891pt}{0.350pt}}
\put(1024,232){\rule[-0.175pt]{1.114pt}{0.350pt}}
\put(1028,231){\rule[-0.175pt]{1.114pt}{0.350pt}}
\put(1033,230){\rule[-0.175pt]{1.114pt}{0.350pt}}
\put(1037,229){\rule[-0.175pt]{1.114pt}{0.350pt}}
\put(1042,228){\rule[-0.175pt]{1.114pt}{0.350pt}}
\put(1047,227){\rule[-0.175pt]{1.114pt}{0.350pt}}
\put(1051,226){\rule[-0.175pt]{1.114pt}{0.350pt}}
\put(1056,225){\rule[-0.175pt]{1.114pt}{0.350pt}}
\put(1061,224){\rule[-0.175pt]{1.273pt}{0.350pt}}
\put(1066,223){\rule[-0.175pt]{1.273pt}{0.350pt}}
\put(1071,222){\rule[-0.175pt]{1.273pt}{0.350pt}}
\put(1076,221){\rule[-0.175pt]{1.273pt}{0.350pt}}
\put(1082,220){\rule[-0.175pt]{1.273pt}{0.350pt}}
\put(1087,219){\rule[-0.175pt]{1.273pt}{0.350pt}}
\put(1092,218){\rule[-0.175pt]{1.273pt}{0.350pt}}
\put(1098,217){\rule[-0.175pt]{1.486pt}{0.350pt}}
\put(1104,216){\rule[-0.175pt]{1.486pt}{0.350pt}}
\put(1110,215){\rule[-0.175pt]{1.486pt}{0.350pt}}
\put(1116,214){\rule[-0.175pt]{1.486pt}{0.350pt}}
\put(1122,213){\rule[-0.175pt]{1.486pt}{0.350pt}}
\put(1128,212){\rule[-0.175pt]{1.486pt}{0.350pt}}
\put(1134,211){\rule[-0.175pt]{1.831pt}{0.350pt}}
\put(1142,210){\rule[-0.175pt]{1.831pt}{0.350pt}}
\put(1150,209){\rule[-0.175pt]{1.831pt}{0.350pt}}
\put(1157,208){\rule[-0.175pt]{1.831pt}{0.350pt}}
\put(1165,207){\rule[-0.175pt]{1.831pt}{0.350pt}}
\put(1172,206){\rule[-0.175pt]{2.228pt}{0.350pt}}
\put(1182,205){\rule[-0.175pt]{2.228pt}{0.350pt}}
\put(1191,204){\rule[-0.175pt]{2.228pt}{0.350pt}}
\put(1200,203){\rule[-0.175pt]{2.228pt}{0.350pt}}
\put(1210,202){\rule[-0.175pt]{2.971pt}{0.350pt}}
\put(1222,201){\rule[-0.175pt]{2.971pt}{0.350pt}}
\put(1234,200){\rule[-0.175pt]{2.971pt}{0.350pt}}
\put(1247,199){\rule[-0.175pt]{2.971pt}{0.350pt}}
\put(1259,198){\rule[-0.175pt]{2.971pt}{0.350pt}}
\put(1271,197){\rule[-0.175pt]{2.971pt}{0.350pt}}
\put(1284,196){\rule[-0.175pt]{2.971pt}{0.350pt}}
\put(1296,195){\rule[-0.175pt]{2.971pt}{0.350pt}}
\put(1308,194){\rule[-0.175pt]{2.971pt}{0.350pt}}
\put(1321,193){\rule[-0.175pt]{4.457pt}{0.350pt}}
\put(1339,192){\rule[-0.175pt]{4.457pt}{0.350pt}}
\put(1358,191){\rule[-0.175pt]{9.154pt}{0.350pt}}
\put(1396,190){\rule[-0.175pt]{4.457pt}{0.350pt}}
\put(1414,189){\rule[-0.175pt]{4.457pt}{0.350pt}}
\put(1433,188){\rule[-0.175pt]{8.913pt}{0.350pt}}
\put(1470,187){\rule[-0.175pt]{17.827pt}{0.350pt}}
\put(1544,186){\rule[-0.175pt]{26.981pt}{0.350pt}}
\sbox{\plotpoint}{\rule[-0.350pt]{0.700pt}{0.700pt}}%
\put(1606,977){\makebox(0,0)[r]{light higgs exp}}
\put(1628,977){\rule[-0.350pt]{15.899pt}{0.700pt}}
\put(271,582){\usebox{\plotpoint}}
\put(271,582){\usebox{\plotpoint}}
\put(272,581){\usebox{\plotpoint}}
\put(273,580){\usebox{\plotpoint}}
\put(274,579){\usebox{\plotpoint}}
\put(275,578){\usebox{\plotpoint}}
\put(276,577){\usebox{\plotpoint}}
\put(277,576){\usebox{\plotpoint}}
\put(278,575){\usebox{\plotpoint}}
\put(279,574){\usebox{\plotpoint}}
\put(280,573){\usebox{\plotpoint}}
\put(281,572){\usebox{\plotpoint}}
\put(282,571){\usebox{\plotpoint}}
\put(283,570){\usebox{\plotpoint}}
\put(284,569){\usebox{\plotpoint}}
\put(285,568){\usebox{\plotpoint}}
\put(286,567){\usebox{\plotpoint}}
\put(287,566){\usebox{\plotpoint}}
\put(288,565){\usebox{\plotpoint}}
\put(289,564){\usebox{\plotpoint}}
\put(290,563){\usebox{\plotpoint}}
\put(291,562){\usebox{\plotpoint}}
\put(292,561){\usebox{\plotpoint}}
\put(293,560){\usebox{\plotpoint}}
\put(294,559){\usebox{\plotpoint}}
\put(295,558){\usebox{\plotpoint}}
\put(296,557){\usebox{\plotpoint}}
\put(297,556){\usebox{\plotpoint}}
\put(298,555){\usebox{\plotpoint}}
\put(299,554){\usebox{\plotpoint}}
\put(300,553){\usebox{\plotpoint}}
\put(301,552){\usebox{\plotpoint}}
\put(302,551){\usebox{\plotpoint}}
\put(304,550){\usebox{\plotpoint}}
\put(305,549){\usebox{\plotpoint}}
\put(306,548){\usebox{\plotpoint}}
\put(307,547){\usebox{\plotpoint}}
\put(308,546){\usebox{\plotpoint}}
\put(309,545){\usebox{\plotpoint}}
\put(310,544){\usebox{\plotpoint}}
\put(311,543){\usebox{\plotpoint}}
\put(312,542){\usebox{\plotpoint}}
\put(313,541){\usebox{\plotpoint}}
\put(314,540){\usebox{\plotpoint}}
\put(316,539){\usebox{\plotpoint}}
\put(317,538){\usebox{\plotpoint}}
\put(318,537){\usebox{\plotpoint}}
\put(319,536){\usebox{\plotpoint}}
\put(320,535){\usebox{\plotpoint}}
\put(321,534){\usebox{\plotpoint}}
\put(322,533){\usebox{\plotpoint}}
\put(323,532){\usebox{\plotpoint}}
\put(324,531){\usebox{\plotpoint}}
\put(325,530){\usebox{\plotpoint}}
\put(327,529){\usebox{\plotpoint}}
\put(328,528){\usebox{\plotpoint}}
\put(329,527){\usebox{\plotpoint}}
\put(330,526){\usebox{\plotpoint}}
\put(331,525){\usebox{\plotpoint}}
\put(333,524){\usebox{\plotpoint}}
\put(334,523){\usebox{\plotpoint}}
\put(335,522){\usebox{\plotpoint}}
\put(336,521){\usebox{\plotpoint}}
\put(338,520){\usebox{\plotpoint}}
\put(339,519){\usebox{\plotpoint}}
\put(340,518){\usebox{\plotpoint}}
\put(341,517){\usebox{\plotpoint}}
\put(342,516){\usebox{\plotpoint}}
\put(344,515){\usebox{\plotpoint}}
\put(345,514){\usebox{\plotpoint}}
\put(346,513){\usebox{\plotpoint}}
\put(347,512){\usebox{\plotpoint}}
\put(349,511){\usebox{\plotpoint}}
\put(350,510){\usebox{\plotpoint}}
\put(351,509){\usebox{\plotpoint}}
\put(352,508){\usebox{\plotpoint}}
\put(353,507){\usebox{\plotpoint}}
\put(354,506){\usebox{\plotpoint}}
\put(355,505){\usebox{\plotpoint}}
\put(356,504){\usebox{\plotpoint}}
\put(357,503){\usebox{\plotpoint}}
\put(358,502){\usebox{\plotpoint}}
\put(360,501){\usebox{\plotpoint}}
\put(361,500){\usebox{\plotpoint}}
\put(362,499){\usebox{\plotpoint}}
\put(364,498){\usebox{\plotpoint}}
\put(365,497){\usebox{\plotpoint}}
\put(366,496){\usebox{\plotpoint}}
\put(368,495){\usebox{\plotpoint}}
\put(369,494){\usebox{\plotpoint}}
\put(371,493){\usebox{\plotpoint}}
\put(372,492){\usebox{\plotpoint}}
\put(373,491){\usebox{\plotpoint}}
\put(374,490){\usebox{\plotpoint}}
\put(375,489){\usebox{\plotpoint}}
\put(377,488){\usebox{\plotpoint}}
\put(378,487){\usebox{\plotpoint}}
\put(379,486){\usebox{\plotpoint}}
\put(380,485){\usebox{\plotpoint}}
\put(382,484){\usebox{\plotpoint}}
\put(383,483){\usebox{\plotpoint}}
\put(384,482){\usebox{\plotpoint}}
\put(386,481){\usebox{\plotpoint}}
\put(387,480){\usebox{\plotpoint}}
\put(388,479){\usebox{\plotpoint}}
\put(390,478){\usebox{\plotpoint}}
\put(391,477){\usebox{\plotpoint}}
\put(393,476){\usebox{\plotpoint}}
\put(394,475){\usebox{\plotpoint}}
\put(395,474){\usebox{\plotpoint}}
\put(397,473){\usebox{\plotpoint}}
\put(398,472){\usebox{\plotpoint}}
\put(399,471){\usebox{\plotpoint}}
\put(401,470){\usebox{\plotpoint}}
\put(402,469){\usebox{\plotpoint}}
\put(404,468){\usebox{\plotpoint}}
\put(405,467){\usebox{\plotpoint}}
\put(406,466){\usebox{\plotpoint}}
\put(408,465){\usebox{\plotpoint}}
\put(409,464){\usebox{\plotpoint}}
\put(410,463){\usebox{\plotpoint}}
\put(412,462){\usebox{\plotpoint}}
\put(413,461){\usebox{\plotpoint}}
\put(415,460){\usebox{\plotpoint}}
\put(416,459){\usebox{\plotpoint}}
\put(418,458){\usebox{\plotpoint}}
\put(419,457){\usebox{\plotpoint}}
\put(421,456){\usebox{\plotpoint}}
\put(422,455){\usebox{\plotpoint}}
\put(424,454){\usebox{\plotpoint}}
\put(426,453){\usebox{\plotpoint}}
\put(427,452){\usebox{\plotpoint}}
\put(429,451){\usebox{\plotpoint}}
\put(430,450){\usebox{\plotpoint}}
\put(432,449){\usebox{\plotpoint}}
\put(433,448){\usebox{\plotpoint}}
\put(435,447){\usebox{\plotpoint}}
\put(437,446){\usebox{\plotpoint}}
\put(438,445){\usebox{\plotpoint}}
\put(440,444){\usebox{\plotpoint}}
\put(442,443){\usebox{\plotpoint}}
\put(444,442){\usebox{\plotpoint}}
\put(446,441){\usebox{\plotpoint}}
\put(448,440){\usebox{\plotpoint}}
\put(449,439){\usebox{\plotpoint}}
\put(451,438){\usebox{\plotpoint}}
\put(453,437){\usebox{\plotpoint}}
\put(455,436){\usebox{\plotpoint}}
\put(457,435){\usebox{\plotpoint}}
\put(459,434){\usebox{\plotpoint}}
\put(460,433){\usebox{\plotpoint}}
\put(462,432){\usebox{\plotpoint}}
\put(464,431){\usebox{\plotpoint}}
\put(466,430){\usebox{\plotpoint}}
\put(468,429){\usebox{\plotpoint}}
\put(470,428){\usebox{\plotpoint}}
\put(471,427){\usebox{\plotpoint}}
\put(473,426){\usebox{\plotpoint}}
\put(475,425){\usebox{\plotpoint}}
\put(477,424){\usebox{\plotpoint}}
\put(479,423){\usebox{\plotpoint}}
\put(481,422){\usebox{\plotpoint}}
\put(483,421){\usebox{\plotpoint}}
\put(486,420){\usebox{\plotpoint}}
\put(489,419){\usebox{\plotpoint}}
\put(492,418){\usebox{\plotpoint}}
\put(494,417){\usebox{\plotpoint}}
\put(496,416){\usebox{\plotpoint}}
\put(498,415){\usebox{\plotpoint}}
\put(500,414){\usebox{\plotpoint}}
\put(503,413){\usebox{\plotpoint}}
\put(505,412){\usebox{\plotpoint}}
\put(508,411){\usebox{\plotpoint}}
\put(511,410){\usebox{\plotpoint}}
\put(514,409){\rule[-0.350pt]{0.883pt}{0.700pt}}
\put(517,408){\rule[-0.350pt]{0.883pt}{0.700pt}}
\put(521,407){\rule[-0.350pt]{0.883pt}{0.700pt}}
\put(525,406){\rule[-0.350pt]{0.883pt}{0.700pt}}
\put(528,405){\rule[-0.350pt]{0.883pt}{0.700pt}}
\put(532,404){\rule[-0.350pt]{0.883pt}{0.700pt}}
\put(536,403){\rule[-0.350pt]{0.883pt}{0.700pt}}
\put(539,402){\rule[-0.350pt]{0.883pt}{0.700pt}}
\put(543,401){\rule[-0.350pt]{0.883pt}{0.700pt}}
\put(547,400){\rule[-0.350pt]{1.325pt}{0.700pt}}
\put(552,399){\rule[-0.350pt]{1.325pt}{0.700pt}}
\put(558,398){\rule[-0.350pt]{2.650pt}{0.700pt}}
\sbox{\plotpoint}{\rule[-0.500pt]{1.000pt}{1.000pt}}%
\put(1606,932){\makebox(0,0)[r]{interpolation}}
\put(1628,932){\rule[-0.500pt]{15.899pt}{1.000pt}}
\put(271,582){\usebox{\plotpoint}}
\put(271,582){\usebox{\plotpoint}}
\put(272,581){\usebox{\plotpoint}}
\put(273,580){\usebox{\plotpoint}}
\put(274,579){\usebox{\plotpoint}}
\put(275,578){\usebox{\plotpoint}}
\put(276,577){\usebox{\plotpoint}}
\put(277,576){\usebox{\plotpoint}}
\put(278,575){\usebox{\plotpoint}}
\put(279,574){\usebox{\plotpoint}}
\put(280,573){\usebox{\plotpoint}}
\put(281,572){\usebox{\plotpoint}}
\put(282,571){\usebox{\plotpoint}}
\put(283,570){\usebox{\plotpoint}}
\put(284,569){\usebox{\plotpoint}}
\put(285,568){\usebox{\plotpoint}}
\put(286,567){\usebox{\plotpoint}}
\put(287,566){\usebox{\plotpoint}}
\put(288,565){\usebox{\plotpoint}}
\put(289,564){\usebox{\plotpoint}}
\put(291,563){\usebox{\plotpoint}}
\put(292,562){\usebox{\plotpoint}}
\put(293,561){\usebox{\plotpoint}}
\put(294,560){\usebox{\plotpoint}}
\put(295,559){\usebox{\plotpoint}}
\put(296,558){\usebox{\plotpoint}}
\put(297,557){\usebox{\plotpoint}}
\put(298,556){\usebox{\plotpoint}}
\put(299,555){\usebox{\plotpoint}}
\put(300,554){\usebox{\plotpoint}}
\put(301,553){\usebox{\plotpoint}}
\put(302,552){\usebox{\plotpoint}}
\put(303,551){\usebox{\plotpoint}}
\put(304,550){\usebox{\plotpoint}}
\put(305,549){\usebox{\plotpoint}}
\put(306,548){\usebox{\plotpoint}}
\put(307,547){\usebox{\plotpoint}}
\put(308,546){\usebox{\plotpoint}}
\put(309,545){\usebox{\plotpoint}}
\put(310,544){\usebox{\plotpoint}}
\put(311,543){\usebox{\plotpoint}}
\put(312,542){\usebox{\plotpoint}}
\put(313,541){\usebox{\plotpoint}}
\put(315,540){\usebox{\plotpoint}}
\put(316,539){\usebox{\plotpoint}}
\put(317,538){\usebox{\plotpoint}}
\put(318,537){\usebox{\plotpoint}}
\put(319,536){\usebox{\plotpoint}}
\put(321,535){\usebox{\plotpoint}}
\put(322,534){\usebox{\plotpoint}}
\put(323,533){\usebox{\plotpoint}}
\put(324,532){\usebox{\plotpoint}}
\put(325,531){\usebox{\plotpoint}}
\put(327,530){\usebox{\plotpoint}}
\put(328,529){\usebox{\plotpoint}}
\put(329,528){\usebox{\plotpoint}}
\put(330,527){\usebox{\plotpoint}}
\put(331,526){\usebox{\plotpoint}}
\put(332,525){\usebox{\plotpoint}}
\put(333,524){\usebox{\plotpoint}}
\put(334,523){\usebox{\plotpoint}}
\put(336,522){\usebox{\plotpoint}}
\put(337,521){\usebox{\plotpoint}}
\put(338,520){\usebox{\plotpoint}}
\put(339,519){\usebox{\plotpoint}}
\put(340,518){\usebox{\plotpoint}}
\put(341,517){\usebox{\plotpoint}}
\put(342,516){\usebox{\plotpoint}}
\put(343,515){\usebox{\plotpoint}}
\put(345,514){\usebox{\plotpoint}}
\put(346,513){\usebox{\plotpoint}}
\put(347,512){\usebox{\plotpoint}}
\put(348,511){\usebox{\plotpoint}}
\put(349,510){\usebox{\plotpoint}}
\put(351,509){\usebox{\plotpoint}}
\put(352,508){\usebox{\plotpoint}}
\put(353,507){\usebox{\plotpoint}}
\put(354,506){\usebox{\plotpoint}}
\put(355,505){\usebox{\plotpoint}}
\put(357,504){\usebox{\plotpoint}}
\put(358,503){\usebox{\plotpoint}}
\put(359,502){\usebox{\plotpoint}}
\put(360,501){\usebox{\plotpoint}}
\put(361,500){\usebox{\plotpoint}}
\put(363,499){\usebox{\plotpoint}}
\put(364,498){\usebox{\plotpoint}}
\put(365,497){\usebox{\plotpoint}}
\put(366,496){\usebox{\plotpoint}}
\put(368,495){\usebox{\plotpoint}}
\put(369,494){\usebox{\plotpoint}}
\put(370,493){\usebox{\plotpoint}}
\put(371,492){\usebox{\plotpoint}}
\put(373,491){\usebox{\plotpoint}}
\put(374,490){\usebox{\plotpoint}}
\put(375,489){\usebox{\plotpoint}}
\put(376,488){\usebox{\plotpoint}}
\put(378,487){\usebox{\plotpoint}}
\put(379,486){\usebox{\plotpoint}}
\put(380,485){\usebox{\plotpoint}}
\put(381,484){\usebox{\plotpoint}}
\put(383,483){\usebox{\plotpoint}}
\put(384,482){\usebox{\plotpoint}}
\put(386,481){\usebox{\plotpoint}}
\put(387,480){\usebox{\plotpoint}}
\put(388,479){\usebox{\plotpoint}}
\put(390,478){\usebox{\plotpoint}}
\put(391,477){\usebox{\plotpoint}}
\put(393,476){\usebox{\plotpoint}}
\put(394,475){\usebox{\plotpoint}}
\put(395,474){\usebox{\plotpoint}}
\put(397,473){\usebox{\plotpoint}}
\put(398,472){\usebox{\plotpoint}}
\put(399,471){\usebox{\plotpoint}}
\put(401,470){\usebox{\plotpoint}}
\put(402,469){\usebox{\plotpoint}}
\put(404,468){\usebox{\plotpoint}}
\put(405,467){\usebox{\plotpoint}}
\put(406,466){\usebox{\plotpoint}}
\put(408,465){\usebox{\plotpoint}}
\put(409,464){\usebox{\plotpoint}}
\put(410,463){\usebox{\plotpoint}}
\put(412,462){\usebox{\plotpoint}}
\put(413,461){\usebox{\plotpoint}}
\put(414,460){\usebox{\plotpoint}}
\put(416,459){\usebox{\plotpoint}}
\put(417,458){\usebox{\plotpoint}}
\put(419,457){\usebox{\plotpoint}}
\put(420,456){\usebox{\plotpoint}}
\put(422,455){\usebox{\plotpoint}}
\put(423,454){\usebox{\plotpoint}}
\put(425,453){\usebox{\plotpoint}}
\put(426,452){\usebox{\plotpoint}}
\put(428,451){\usebox{\plotpoint}}
\put(429,450){\usebox{\plotpoint}}
\put(431,449){\usebox{\plotpoint}}
\put(432,448){\usebox{\plotpoint}}
\put(434,447){\usebox{\plotpoint}}
\put(435,446){\usebox{\plotpoint}}
\put(437,445){\usebox{\plotpoint}}
\put(438,444){\usebox{\plotpoint}}
\put(440,443){\usebox{\plotpoint}}
\put(441,442){\usebox{\plotpoint}}
\put(443,441){\usebox{\plotpoint}}
\put(444,440){\usebox{\plotpoint}}
\put(446,439){\usebox{\plotpoint}}
\put(447,438){\usebox{\plotpoint}}
\put(449,437){\usebox{\plotpoint}}
\put(450,436){\usebox{\plotpoint}}
\put(452,435){\usebox{\plotpoint}}
\put(453,434){\usebox{\plotpoint}}
\put(455,433){\usebox{\plotpoint}}
\put(456,432){\usebox{\plotpoint}}
\put(458,431){\usebox{\plotpoint}}
\put(459,430){\usebox{\plotpoint}}
\put(461,429){\usebox{\plotpoint}}
\put(462,428){\usebox{\plotpoint}}
\put(464,427){\usebox{\plotpoint}}
\put(466,426){\usebox{\plotpoint}}
\put(467,425){\usebox{\plotpoint}}
\put(469,424){\usebox{\plotpoint}}
\put(470,423){\usebox{\plotpoint}}
\put(472,422){\usebox{\plotpoint}}
\put(474,421){\usebox{\plotpoint}}
\put(475,420){\usebox{\plotpoint}}
\put(477,419){\usebox{\plotpoint}}
\put(478,418){\usebox{\plotpoint}}
\put(480,417){\usebox{\plotpoint}}
\put(482,416){\usebox{\plotpoint}}
\put(483,415){\usebox{\plotpoint}}
\put(485,414){\usebox{\plotpoint}}
\put(487,413){\usebox{\plotpoint}}
\put(488,412){\usebox{\plotpoint}}
\put(490,411){\usebox{\plotpoint}}
\put(491,410){\usebox{\plotpoint}}
\put(493,409){\usebox{\plotpoint}}
\put(495,408){\usebox{\plotpoint}}
\put(497,407){\usebox{\plotpoint}}
\put(499,406){\usebox{\plotpoint}}
\put(501,405){\usebox{\plotpoint}}
\put(503,404){\usebox{\plotpoint}}
\put(505,403){\usebox{\plotpoint}}
\put(507,402){\usebox{\plotpoint}}
\put(509,401){\usebox{\plotpoint}}
\put(510,400){\usebox{\plotpoint}}
\put(512,399){\usebox{\plotpoint}}
\put(514,398){\usebox{\plotpoint}}
\put(515,397){\usebox{\plotpoint}}
\put(517,396){\usebox{\plotpoint}}
\put(519,395){\usebox{\plotpoint}}
\put(520,394){\usebox{\plotpoint}}
\put(522,393){\usebox{\plotpoint}}
\put(524,392){\usebox{\plotpoint}}
\put(525,391){\usebox{\plotpoint}}
\put(527,390){\usebox{\plotpoint}}
\put(528,389){\usebox{\plotpoint}}
\put(531,388){\usebox{\plotpoint}}
\put(533,387){\usebox{\plotpoint}}
\put(535,386){\usebox{\plotpoint}}
\put(537,385){\usebox{\plotpoint}}
\put(539,384){\usebox{\plotpoint}}
\put(541,383){\usebox{\plotpoint}}
\put(543,382){\usebox{\plotpoint}}
\put(545,381){\usebox{\plotpoint}}
\put(547,380){\usebox{\plotpoint}}
\put(549,379){\usebox{\plotpoint}}
\put(551,378){\usebox{\plotpoint}}
\put(553,377){\usebox{\plotpoint}}
\put(555,376){\usebox{\plotpoint}}
\put(557,375){\usebox{\plotpoint}}
\put(559,374){\usebox{\plotpoint}}
\put(561,373){\usebox{\plotpoint}}
\put(563,372){\usebox{\plotpoint}}
\put(565,371){\usebox{\plotpoint}}
\put(568,370){\usebox{\plotpoint}}
\put(570,369){\usebox{\plotpoint}}
\put(572,368){\usebox{\plotpoint}}
\put(574,367){\usebox{\plotpoint}}
\put(576,366){\usebox{\plotpoint}}
\put(578,365){\usebox{\plotpoint}}
\put(580,364){\usebox{\plotpoint}}
\put(582,363){\usebox{\plotpoint}}
\put(584,362){\usebox{\plotpoint}}
\put(586,361){\usebox{\plotpoint}}
\put(588,360){\usebox{\plotpoint}}
\put(590,359){\usebox{\plotpoint}}
\put(592,358){\usebox{\plotpoint}}
\put(594,357){\usebox{\plotpoint}}
\put(596,356){\usebox{\plotpoint}}
\put(598,355){\usebox{\plotpoint}}
\put(600,354){\usebox{\plotpoint}}
\put(602,353){\usebox{\plotpoint}}
\put(605,352){\usebox{\plotpoint}}
\put(607,351){\usebox{\plotpoint}}
\put(609,350){\usebox{\plotpoint}}
\put(612,349){\usebox{\plotpoint}}
\put(614,348){\usebox{\plotpoint}}
\put(616,347){\usebox{\plotpoint}}
\put(618,346){\usebox{\plotpoint}}
\put(621,345){\usebox{\plotpoint}}
\put(623,344){\usebox{\plotpoint}}
\put(626,343){\usebox{\plotpoint}}
\put(628,342){\usebox{\plotpoint}}
\put(631,341){\usebox{\plotpoint}}
\put(633,340){\usebox{\plotpoint}}
\put(636,339){\usebox{\plotpoint}}
\put(638,338){\usebox{\plotpoint}}
\put(641,337){\usebox{\plotpoint}}
\put(643,336){\usebox{\plotpoint}}
\put(646,335){\usebox{\plotpoint}}
\put(648,334){\usebox{\plotpoint}}
\put(650,333){\usebox{\plotpoint}}
\put(653,332){\usebox{\plotpoint}}
\put(655,331){\usebox{\plotpoint}}
\put(658,330){\usebox{\plotpoint}}
\put(660,329){\usebox{\plotpoint}}
\put(663,328){\usebox{\plotpoint}}
\put(665,327){\usebox{\plotpoint}}
\put(668,326){\usebox{\plotpoint}}
\put(670,325){\usebox{\plotpoint}}
\put(673,324){\usebox{\plotpoint}}
\put(675,323){\usebox{\plotpoint}}
\put(678,322){\usebox{\plotpoint}}
\put(681,321){\usebox{\plotpoint}}
\put(684,320){\usebox{\plotpoint}}
\put(686,319){\usebox{\plotpoint}}
\put(689,318){\usebox{\plotpoint}}
\put(692,317){\usebox{\plotpoint}}
\put(695,316){\usebox{\plotpoint}}
\put(698,315){\usebox{\plotpoint}}
\put(701,314){\usebox{\plotpoint}}
\put(704,313){\usebox{\plotpoint}}
\put(707,312){\usebox{\plotpoint}}
\put(710,311){\usebox{\plotpoint}}
\put(713,310){\usebox{\plotpoint}}
\put(716,309){\usebox{\plotpoint}}
\put(719,308){\usebox{\plotpoint}}
\put(722,307){\usebox{\plotpoint}}
\put(725,306){\usebox{\plotpoint}}
\put(728,305){\usebox{\plotpoint}}
\put(731,304){\usebox{\plotpoint}}
\put(734,303){\usebox{\plotpoint}}
\put(737,302){\usebox{\plotpoint}}
\put(740,301){\usebox{\plotpoint}}
\put(743,300){\usebox{\plotpoint}}
\put(746,299){\usebox{\plotpoint}}
\put(750,298){\usebox{\plotpoint}}
\put(753,297){\usebox{\plotpoint}}
\put(756,296){\usebox{\plotpoint}}
\put(759,295){\usebox{\plotpoint}}
\put(762,294){\usebox{\plotpoint}}
\put(765,293){\usebox{\plotpoint}}
\put(768,292){\usebox{\plotpoint}}
\put(771,291){\usebox{\plotpoint}}
\put(775,290){\usebox{\plotpoint}}
\put(779,289){\usebox{\plotpoint}}
\put(783,288){\usebox{\plotpoint}}
\put(786,287){\usebox{\plotpoint}}
\put(790,286){\usebox{\plotpoint}}
\put(793,285){\usebox{\plotpoint}}
\put(796,284){\usebox{\plotpoint}}
\put(799,283){\usebox{\plotpoint}}
\put(802,282){\usebox{\plotpoint}}
\put(805,281){\usebox{\plotpoint}}
\put(808,280){\usebox{\plotpoint}}
\put(812,279){\usebox{\plotpoint}}
\put(815,278){\usebox{\plotpoint}}
\put(819,277){\usebox{\plotpoint}}
\put(822,276){\rule[-0.500pt]{1.144pt}{1.000pt}}
\put(827,275){\rule[-0.500pt]{1.144pt}{1.000pt}}
\put(832,274){\rule[-0.500pt]{1.144pt}{1.000pt}}
\put(837,273){\rule[-0.500pt]{1.144pt}{1.000pt}}
\put(842,272){\usebox{\plotpoint}}
\put(845,271){\usebox{\plotpoint}}
\put(849,270){\usebox{\plotpoint}}
\put(852,269){\usebox{\plotpoint}}
\put(856,268){\usebox{\plotpoint}}
\put(859,267){\usebox{\plotpoint}}
\put(863,266){\usebox{\plotpoint}}
\put(867,265){\usebox{\plotpoint}}
\put(871,264){\usebox{\plotpoint}}
\put(875,263){\usebox{\plotpoint}}
\put(878,262){\rule[-0.500pt]{1.084pt}{1.000pt}}
\put(883,261){\rule[-0.500pt]{1.084pt}{1.000pt}}
\put(888,260){\rule[-0.500pt]{1.084pt}{1.000pt}}
\put(892,259){\rule[-0.500pt]{1.084pt}{1.000pt}}
\put(897,258){\rule[-0.500pt]{1.084pt}{1.000pt}}
\put(901,257){\rule[-0.500pt]{1.084pt}{1.000pt}}
\put(906,256){\rule[-0.500pt]{1.084pt}{1.000pt}}
\put(910,255){\rule[-0.500pt]{1.084pt}{1.000pt}}
\put(915,254){\rule[-0.500pt]{1.144pt}{1.000pt}}
\put(919,253){\rule[-0.500pt]{1.144pt}{1.000pt}}
\put(924,252){\rule[-0.500pt]{1.144pt}{1.000pt}}
\put(929,251){\rule[-0.500pt]{1.144pt}{1.000pt}}
\put(934,250){\rule[-0.500pt]{1.084pt}{1.000pt}}
\put(938,249){\rule[-0.500pt]{1.084pt}{1.000pt}}
\put(943,248){\rule[-0.500pt]{1.084pt}{1.000pt}}
\put(947,247){\rule[-0.500pt]{1.084pt}{1.000pt}}
\put(952,246){\rule[-0.500pt]{1.526pt}{1.000pt}}
\put(958,245){\rule[-0.500pt]{1.526pt}{1.000pt}}
\put(964,244){\rule[-0.500pt]{1.526pt}{1.000pt}}
\put(970,243){\rule[-0.500pt]{1.084pt}{1.000pt}}
\put(975,242){\rule[-0.500pt]{1.084pt}{1.000pt}}
\put(980,241){\rule[-0.500pt]{1.084pt}{1.000pt}}
\put(984,240){\rule[-0.500pt]{1.084pt}{1.000pt}}
\put(989,239){\rule[-0.500pt]{1.445pt}{1.000pt}}
\put(995,238){\rule[-0.500pt]{1.445pt}{1.000pt}}
\put(1001,237){\rule[-0.500pt]{1.445pt}{1.000pt}}
\put(1007,236){\rule[-0.500pt]{1.144pt}{1.000pt}}
\put(1011,235){\rule[-0.500pt]{1.144pt}{1.000pt}}
\put(1016,234){\rule[-0.500pt]{1.144pt}{1.000pt}}
\put(1021,233){\rule[-0.500pt]{1.144pt}{1.000pt}}
\put(1026,232){\rule[-0.500pt]{1.445pt}{1.000pt}}
\put(1032,231){\rule[-0.500pt]{1.445pt}{1.000pt}}
\put(1038,230){\rule[-0.500pt]{1.445pt}{1.000pt}}
\put(1044,229){\rule[-0.500pt]{1.526pt}{1.000pt}}
\put(1050,228){\rule[-0.500pt]{1.526pt}{1.000pt}}
\put(1056,227){\rule[-0.500pt]{1.526pt}{1.000pt}}
\put(1063,226){\rule[-0.500pt]{1.445pt}{1.000pt}}
\put(1069,225){\rule[-0.500pt]{1.445pt}{1.000pt}}
\put(1075,224){\rule[-0.500pt]{1.445pt}{1.000pt}}
\put(1081,223){\rule[-0.500pt]{1.445pt}{1.000pt}}
\put(1087,222){\rule[-0.500pt]{1.445pt}{1.000pt}}
\put(1093,221){\rule[-0.500pt]{1.445pt}{1.000pt}}
\put(1099,220){\rule[-0.500pt]{2.289pt}{1.000pt}}
\put(1108,219){\rule[-0.500pt]{2.289pt}{1.000pt}}
\put(1118,218){\rule[-0.500pt]{1.445pt}{1.000pt}}
\put(1124,217){\rule[-0.500pt]{1.445pt}{1.000pt}}
\put(1130,216){\rule[-0.500pt]{1.445pt}{1.000pt}}
\put(1136,215){\rule[-0.500pt]{1.526pt}{1.000pt}}
\put(1142,214){\rule[-0.500pt]{1.526pt}{1.000pt}}
\put(1148,213){\rule[-0.500pt]{1.526pt}{1.000pt}}
\put(1155,212){\rule[-0.500pt]{2.168pt}{1.000pt}}
\put(1164,211){\rule[-0.500pt]{2.168pt}{1.000pt}}
\put(1173,210){\rule[-0.500pt]{2.168pt}{1.000pt}}
\put(1182,209){\rule[-0.500pt]{2.168pt}{1.000pt}}
\put(1191,208){\rule[-0.500pt]{1.526pt}{1.000pt}}
\put(1197,207){\rule[-0.500pt]{1.526pt}{1.000pt}}
\put(1203,206){\rule[-0.500pt]{1.526pt}{1.000pt}}
\put(1210,205){\rule[-0.500pt]{2.168pt}{1.000pt}}
\put(1219,204){\rule[-0.500pt]{2.168pt}{1.000pt}}
\put(1228,203){\rule[-0.500pt]{2.289pt}{1.000pt}}
\put(1237,202){\rule[-0.500pt]{2.289pt}{1.000pt}}
\put(1247,201){\rule[-0.500pt]{2.168pt}{1.000pt}}
\put(1256,200){\rule[-0.500pt]{2.168pt}{1.000pt}}
\put(1265,199){\rule[-0.500pt]{2.168pt}{1.000pt}}
\put(1274,198){\rule[-0.500pt]{2.168pt}{1.000pt}}
\put(1283,197){\rule[-0.500pt]{2.289pt}{1.000pt}}
\put(1292,196){\rule[-0.500pt]{2.289pt}{1.000pt}}
\put(1302,195){\rule[-0.500pt]{4.336pt}{1.000pt}}
\put(1320,194){\rule[-0.500pt]{2.289pt}{1.000pt}}
\put(1329,193){\rule[-0.500pt]{2.289pt}{1.000pt}}
\put(1339,192){\rule[-0.500pt]{4.336pt}{1.000pt}}
\put(1357,191){\rule[-0.500pt]{2.168pt}{1.000pt}}
\put(1366,190){\rule[-0.500pt]{2.168pt}{1.000pt}}
\put(1375,189){\rule[-0.500pt]{4.577pt}{1.000pt}}
\put(1394,188){\rule[-0.500pt]{4.336pt}{1.000pt}}
\put(1412,187){\rule[-0.500pt]{4.577pt}{1.000pt}}
\put(1431,186){\rule[-0.500pt]{4.336pt}{1.000pt}}
\put(1449,185){\rule[-0.500pt]{8.913pt}{1.000pt}}
\put(1486,184){\rule[-0.500pt]{22.163pt}{1.000pt}}
\put(1578,185){\rule[-0.500pt]{4.336pt}{1.000pt}}
\put(1596,186){\rule[-0.500pt]{4.577pt}{1.000pt}}
\put(1615,187){\rule[-0.500pt]{4.336pt}{1.000pt}}
\put(1633,188){\rule[-0.500pt]{4.336pt}{1.000pt}}
\sbox{\plotpoint}{\rule[-0.250pt]{0.500pt}{0.500pt}}%
\put(1606,887){\makebox(0,0)[r]{leading $m_t^4$}}
\put(1628,887){\usebox{\plotpoint}}
\put(1648,887){\usebox{\plotpoint}}
\put(1669,887){\usebox{\plotpoint}}
\put(1690,887){\usebox{\plotpoint}}
\put(1694,887){\usebox{\plotpoint}}
\put(268,1040){\usebox{\plotpoint}}
\put(268,1040){\usebox{\plotpoint}}
\put(277,1021){\usebox{\plotpoint}}
\put(289,1004){\usebox{\plotpoint}}
\put(300,987){\usebox{\plotpoint}}
\put(312,970){\usebox{\plotpoint}}
\put(325,953){\usebox{\plotpoint}}
\put(338,937){\usebox{\plotpoint}}
\put(352,922){\usebox{\plotpoint}}
\put(366,906){\usebox{\plotpoint}}
\put(380,891){\usebox{\plotpoint}}
\put(395,876){\usebox{\plotpoint}}
\put(409,862){\usebox{\plotpoint}}
\put(425,848){\usebox{\plotpoint}}
\put(441,835){\usebox{\plotpoint}}
\put(456,821){\usebox{\plotpoint}}
\put(473,808){\usebox{\plotpoint}}
\put(489,796){\usebox{\plotpoint}}
\put(506,783){\usebox{\plotpoint}}
\put(523,772){\usebox{\plotpoint}}
\put(540,760){\usebox{\plotpoint}}
\put(558,748){\usebox{\plotpoint}}
\put(575,738){\usebox{\plotpoint}}
\put(593,727){\usebox{\plotpoint}}
\put(612,718){\usebox{\plotpoint}}
\put(630,708){\usebox{\plotpoint}}
\put(648,697){\usebox{\plotpoint}}
\put(667,688){\usebox{\plotpoint}}
\put(685,679){\usebox{\plotpoint}}
\put(704,671){\usebox{\plotpoint}}
\put(723,663){\usebox{\plotpoint}}
\put(742,654){\usebox{\plotpoint}}
\put(761,646){\usebox{\plotpoint}}
\put(781,639){\usebox{\plotpoint}}
\put(800,631){\usebox{\plotpoint}}
\put(819,624){\usebox{\plotpoint}}
\put(839,617){\usebox{\plotpoint}}
\put(859,610){\usebox{\plotpoint}}
\put(878,604){\usebox{\plotpoint}}
\put(898,598){\usebox{\plotpoint}}
\put(918,591){\usebox{\plotpoint}}
\put(938,585){\usebox{\plotpoint}}
\put(958,580){\usebox{\plotpoint}}
\put(978,574){\usebox{\plotpoint}}
\put(998,569){\usebox{\plotpoint}}
\put(1018,563){\usebox{\plotpoint}}
\put(1038,558){\usebox{\plotpoint}}
\put(1058,554){\usebox{\plotpoint}}
\put(1079,549){\usebox{\plotpoint}}
\put(1099,545){\usebox{\plotpoint}}
\put(1119,540){\usebox{\plotpoint}}
\put(1139,536){\usebox{\plotpoint}}
\put(1160,532){\usebox{\plotpoint}}
\put(1180,528){\usebox{\plotpoint}}
\put(1200,524){\usebox{\plotpoint}}
\put(1221,520){\usebox{\plotpoint}}
\put(1241,517){\usebox{\plotpoint}}
\put(1262,514){\usebox{\plotpoint}}
\put(1282,510){\usebox{\plotpoint}}
\put(1303,507){\usebox{\plotpoint}}
\put(1323,504){\usebox{\plotpoint}}
\put(1344,501){\usebox{\plotpoint}}
\put(1364,498){\usebox{\plotpoint}}
\put(1385,496){\usebox{\plotpoint}}
\put(1405,492){\usebox{\plotpoint}}
\put(1426,490){\usebox{\plotpoint}}
\put(1447,488){\usebox{\plotpoint}}
\put(1467,485){\usebox{\plotpoint}}
\put(1488,482){\usebox{\plotpoint}}
\put(1509,481){\usebox{\plotpoint}}
\put(1529,479){\usebox{\plotpoint}}
\put(1550,477){\usebox{\plotpoint}}
\put(1570,474){\usebox{\plotpoint}}
\put(1591,472){\usebox{\plotpoint}}
\put(1612,471){\usebox{\plotpoint}}
\put(1632,469){\usebox{\plotpoint}}
\put(1653,467){\usebox{\plotpoint}}
\put(1662,467){\usebox{\plotpoint}}
\end{picture}
\caption{\sf $\Delta\hat{\rho}^{(2)}$ in units $N_c (\acur/(4 \pi \scur))^2
 (\mt^2/(4 \mz^2 \ccur))^2$.
 The solid lines 
represent the expressions in Eq.(10) and a simple interpolating function. 
The dotted line represents the leading \gmuq\ term, first two lines of
 Eq.(10b).}
\end{figure}
In Fig.1 we show the behaviour of the two expansions in \eqs{e2.14}
 as functions of $\mh$,
for $\mt=180$GeV and $\scur=0.2314$. 
We also show the leading $\mt^4$ contribution (first two 
lines of \equ{e2.14b}), and a simple interpolation curve which reproduces 
the light and heavy higgs expansions with very good accuracy
in their expected ranges of validity. 
For 160GeV$\,\lequiv\, \mt \,\lequiv\, 200$GeV, $\mh \lequiv 3.8 \, \mt$
and $\scur\approx 0.2314$,
 this interpolation function takes the form:
$f(h,\mt)=  -15.642 +    0.036382\,\mt 
   +\sqrt{h}(2.301 - 0.01343\,\mt) + 
h( 0.01809\,\mt-9.953 ) 
+h^2 (5.687 - 0.01568\,\mt) + 
   h^3 ( 0.005369\,\mt-1.647 )+ 
   h^4 (0.1852 - 0.000646\,\mt)  $\, ($h=\mh/\mt$).
The figure clearly  shows the magnitude of the new $\amtd$ correction:
leaving aside the very low Higgs region, where large cancellations 
conspire to make the leading \amtq\ correction particularly small, we
observe that in most of the allowed Higgs range the $\amtd$ term
 is roughly as large as the leading correction. 
With the purpose of investigating the magnitude of the \amtd\
corrections to physical observables, we have calculated 
the theoretical predictions for $\mw$ and $\sincur$ for different values of
 $\mt$ and $\mh$, solving iteratively Eqs. \ref{e2.4} and \ref{e2.6}.
In Table 1 we report the shifts induced in $\mw$
and $\scur$ by the new \amtd\ contributions 
with respect to the inclusion of the leading \amtq\ term only.
We emphasize that because of the presence of irreducible \amtd\
effects in the $\gamma-Z$ mixing on the Z-peak, 
 the shifts obtained for $\scur$ cannot be simply related \cite{eff} to the 
effective sine measured at LEP and SLC. 

In conclusion, we have presented the 
results of a complete analytic calculation 
of the \amtd\ effects on the interdependence of $\mw$, $\mz$, and $G_\mu$.
We find that for  most of the relevant 
$\mh$ values the \amtd\ correction is of the order of  the leading
\amtq\ term, and that the effect on the prediction for 
$\mw$ from $\alpha$, $G_\mu$, and $\mz$,
can be as large as 20MeV, depending on the top and Higgs masses (see
 Table 1).
Details of this calculation will be presented in a forthcoming communication.

\subsection*{Acknowledgments}
We are happy to thank S. Fanchiotti and F. Feruglio for collaboration 
in the early stage of this  project, and for helpful and interesting 
discussions. We are also grateful to S.~Bauberger,
K.~Chetyrkin, S.~Davidson, B.~Kniehl, G.~Passarino
 A.~Sirlin, and B.~Tausk 
for discussions and helpful communications.
G.D. would like to thank the Theory Group of the Max Planck Institut
f\"ur Physik in Munich for the kind hospitality, and P.G. would like 
to thank the Physics Departments of New York University (where part of the 
work was carried out) and of the University of Padova 
for hospitality and support.

\newpage
\renewcommand{\arraystretch}{1.3}
\begin{table} 
\[
\begin{array}{|c||c||c|c|c|}\hline
\mh & \mt & \delta\mw ({\rm MeV}) & \delta\scur \ (10^{-4})  & 
\delta\Delta\hat{\rho}^{(2)} (10^{-4}) \\  \hline\hline
 & 160 & -10.1 & 0.59 &    -1.72   \\ \hline
65 &180 & -15.2 & 0.88 & -2.57 \\ \hline
   &200 &- 21.2 & 1.23 &-3.62    \\ \hline \hline
 & 160 & -9.6 &0.55  &-1.63       \\ \hline
100 &180 &-14.3 & 0.82  &-2.45   \\ \hline
   &200 &- 20.2 &1.15  & -3.46  \\ \hline \hline
 & 160 &-8.3 & 0.46 &  -1.41     \\ \hline
300 &180 & -12.5 & 0.70 & -2.14   \\ \hline
   &200 &- 17.7 & 0.98 &-3.05    \\ \hline \hline
 & 160 & -7.0 &   0.39& -1.16      \\ \hline
600 &180 &-11.0  & 0.62&-1.87   \\ \hline
   &200 & - 16.2& 0.90 &-2.76    \\ \hline \hline
\end{array}            
\]
\caption{\sf Shifts induced by the complete \amtd\ corrections in $\mw$, 
$\sin^2 \hat{\theta}_{\msbar}(\mz^2)$, and  $\hat{\rho}$ with respect 
to the inclusion of the 
leading \amtq\ term alone.
Top and Higgs masses are expressed in GeV.}
\end{table}
\end{document}